\documentclass[letterpaper,twocolumn,10pt]{article}
\usepackage{usenix}
\usepackage{cite}
\usepackage{amsmath,amssymb,amsfonts}
\usepackage{algorithmic}
\usepackage{graphicx}
\usepackage{textcomp}
\usepackage{xcolor}
\def\BibTeX{{\rm B\kern-.05em{\sc i\kern-.025em b}\kern-.08em
    T\kern-.1667em\lower.7ex\hbox{E}\kern-.125emX}}

\usepackage{tabularx}
\usepackage{makecell}
\usepackage{booktabs}
\usepackage[labelfont=bf]{caption}
\usepackage{xurl}

\usepackage{pifont}  

\usepackage{epsf}
\usepackage{epsfig}
\usepackage{graphicx}
\usepackage{ifthen}
\usepackage{array}
\usepackage{comment}
\usepackage{url}
\usepackage{cellspace}
\usepackage{latexsym}
\usepackage{amsfonts}
\usepackage{amsmath}
\usepackage{multirow}
\usepackage{multicol}
\usepackage{tabularx}
\usepackage{listings}
\usepackage{mathtools}
\usepackage{wrapfig}
\usepackage{xspace}

\usepackage{enumitem}

\usepackage{footnote}
\usepackage{gensymb}
\usepackage{tabularx}
\usepackage{makecell}
\newboolean{showcomments}
\usepackage[font=footnotesize,caption=true]{subfig}

\usepackage[utf8]{inputenc}
\usepackage{tikz}
\usepackage{pgfplots}
\pgfplotsset{width=7.5cm,compat=1.12}
\usepgfplotslibrary{fillbetween}

\usepackage[skins]{tcolorbox}
\newtcolorbox{takeaway}[1]{
	lower separated=false,
	colback=blue!2!white,
	colframe=teal!90!black,
	fonttitle=\sffamily\bfseries,
	left=1pt,
	right=1pt,
	bottom=1pt,
	top=4pt,
	colbacktitle=teal!90!black,
	coltitle=blue!2!white,
	enhanced,
	attach boxed title to top left={yshift=-0.1in,xshift=0.15in},
	boxed title style={boxrule=0pt,colframe=white,},
	title={\color{white}{#1}}
}

\setboolean{showcomments}{true}

\ifthenelse{\boolean{showcomments}}
  {\newcommand{\nb}[2]{
    \fbox{\bfseries\sffamily\scriptsize#1}
    {\sf\small$\blacktriangleright$\textit{#2}$\blacktriangleleft$}
   }
   
  }
  {\newcommand{\nb}[2]{}
   
  }

\definecolor{dkgreen}{rgb}{0,0.6,0}
\definecolor{gray}{rgb}{0.5,0.5,0.5}
\definecolor{mauve}{rgb}{0.58,0,0.82}
\definecolor{shadecolor}{rgb}{0.95,0.95,0.95}
\definecolor{pblue}{rgb}{0.13,0.13,1}
\definecolor{pgreen}{rgb}{0,0.5,0}
\definecolor{pred}{rgb}{0.9,0,0}
\definecolor{pgrey}{rgb}{0.46,0.45,0.48}

\lstdefinestyle{toplisting}{
  float=tp,
  floatplacement=tbp,
}

\newcommand{\ie}{i.e.,\xspace}
\newcommand{\eg}{e.g.,\xspace}
\newcommand{\etc}{\textit{etc.}\xspace}
\newcommand{\etal}{\textit{et al.}\xspace}

\newcommand{\ciphergetinstance}{\texttt{\small Cipher.getInstance(<parameter>)}\xspace}
\newcommand{\cipherget}{\texttt{\small Cipher.getInstance}\xspace}

\newcommand{\des}{\texttt{des}\xspace}
\newcommand{\DES}{\texttt{DES}\xspace}

\newcommand{\trustManager}{\texttt{\small TrustManager}\xspace}

\newcommand{\checkServerTrusted}{\texttt{\small checkServerTrusted}\xspace}

\newcommand{\hostnameVerifier}{\texttt{\small HostnameVerifier}\xspace}

\newcommand{\secretkeyspec}{\texttt{\small SecretKeySpec}\xspace}
\newcommand\code[1]{{\tt \small {#1}}}
\newcommand{\dbq}{\texttt{\char`\"}}

\newcommand{\redcheckmark}{\color{red}\ding{54}}
\newcommand{\graycheckmark}{\color{gray}\ding{54}}

\newcommand{\greencheckmark}{\color{green}\ding{52}}

\usepackage[autostyle]{csquotes}
\MakeOuterQuote{"}

\newcounter{fcounter}
\newcommand{\finding}[1]{\refstepcounter{fcounter}
\vspace{0.25em}\noindent\fbox{%
    \parbox{0.95\linewidth}{%
  \vspace{0.3em}{\bf
  {Finding~\arabic{fcounter}~(\fnumber{\arabic{fcounter}})}~--} {#1}
\vspace{0.3em}
  }
}
\vspace{0.25em}

}
\newcommand\fnumber[1]{{$\mathcal{F}_{#1}$}}
\newcommand\score[1]{{\tt {[SCORE $=$ #1]}}}

\newcommand{\myparagraph}[1]{\vspace{0.3em}\noindent{\bf #1:}}

\setlist[itemize]{leftmargin=*, noitemsep, topsep=1pt}
\setlist[enumerate]{leftmargin=*, noitemsep, topsep=1pt}

\makeatletter
\g@addto@macro\normalsize{%
	\setlength\abovedisplayskip{0pt}
	\setlength\belowdisplayskip{0pt}
	\setlength\abovedisplayshortskip{-10pt}
	\setlength\belowdisplayshortskip{0pt}
}
\makeatother

\DeclareGraphicsExtensions{.pdf,.png}
\graphicspath{{img/}}

\usepackage{xcolor}
\begin{document}

\date{}

\title{\Large \bf {\em From base cases to backdoors:} An Empirical Study of Unnatural Crypto-API Misuse}

\author{
	{\rm Victor Olaiya}\\
	William \& Mary
	\and
	{\rm Adwait Nadkarni}\\
	William \& Mary
} 

\maketitle

\begin{abstract}

Tools focused on cryptographic API misuse often detect the most basic expressions of the vulnerable use, and are unable to detect non-trivial variants.
The question of whether tools should be designed to detect such variants can only be answered if we know {\em how} developers use and misuse cryptographic APIs in the wild, and in particular, {\em what the unnatural usage of such APIs looks like}.
This paper presents the first large-scale study that characterizes unnatural crypto-API usage through a qualitative analysis of $5,704$ representative API invocations.
We develop an intuitive complexity metric to stratify $140,431$ crypto-API invocations obtained from $20,508$ Android applications, allowing us to sample $5,704$ invocations that are representative of all strata, with each stratum consisting of invocations with similar complexity/naturalness.
We qualitatively analyze the $5,704$ sampled invocations using manual reverse engineering, through an in-depth investigation that involves the development of minimal examples and exploration of native code. 
Our study results in two detailed taxonomies of unnatural crypto-API misuse, along with $17$ key findings that show the presence of highly unusual misuse, evasive code, and the inability of popular tools to reason about even mildly unconventional usage.
Our findings lead to four key takeaways that  inform future work focused on detecting unnatural crypto-API misuse.

\end{abstract}

\section{Introduction}
\label{sec:intro}

Cryptography plays a critical role in helping developers protect sensitive data.
However, developers have a history of misusing crypto-APIs~\cite{ami2022crypto,egele2013empirical,fahl2012eve,oltrogge2021eve}, such as by using weak ciphers or foregoing necessary certificate verification, thereby putting user data at risk.
In response, the security community has developed tools such as CryptoGuard~\cite{rahaman2019cryptoguard} and CogniCrypt~\cite{kruger2020cognicryptgen} to detect such crypto-API misuse. 
Recent work by Ami et al. shows that these popular crypto-API misuse detectors, or {\em crypto-detectors} in short, are highly ineffective at detecting even the misuse they claim to detect~\cite{ami2022crypto}.
Particularly, while these tools can detect the simplest expression of a misuse, they often struggle to detect non-trivial {\em variants} of the same.
For example, consider the instantiation of the vulnerable \des cipher using the \ciphergetinstance API of the Java Cryptography Architecture (JCA).
The most basic instance would essentially be \code{Cipher.getInstance(\dbq{}DES\dbq{})}, which most tools would detect. 
However, Ami et al. showed that tools may not detect even a simple variant of the same vulnerable use that specifies the argument in lowercase, \ie as ``\code{des}'', let alone a more complex variant such as in Listing~\ref{lst:desreplace}.  

\noindent
\begin{minipage}{\linewidth}
	\begin{lstlisting}[frame=tb,caption={{\small An unusual use of the \code{DES} cipher.}}, label={lst:desreplace},language=java,basicstyle=\fontencoding{T1}\fontfamily{lmtt}\bfseries\scriptsize]
Cipher.getInstance("DE$S".replace("$", ""));
	\end{lstlisting}
\end{minipage}

Tool designers have argued that such expression of crypto-APIs, as in Listing~\ref{lst:desreplace}, is {\em unnatural}, \ie overly complex and unlikely to be present in real software, and hence, not something tools should be expected to detect~\cite{ami2022crypto}. 
However, a counter-argument can also be made, that such perceived unnaturalness may exist in code in the wild, and hence, tools should look beyond typically conventional usage.
This contention exposes a fundamental knowledge gap in the design space of crypto-API vulnerability detection: {\em we do not know enough about the {\bf \em unnatural, unconventional expression} of crypto-API in the wild}, both in terms of the characteristics of such expression, or its prevalence in software.
Addressing this gap through systematic study of real-world API usage would not only help address this contention, but also motivate the development of targeted techniques to find highly unnatural but relevant misuse.
Therefore, this paper is motivated by a simple but critical {\bf \em research question}: {\em what are the odd, unconventional, ways in which developers (mis)use crypto APIs in the wild?}

\myparagraph{Contributions} This paper is the first to develop a qualitative characterization of the unconventional use and misuse of crypto-API in the wild, through a study of $5,704$ invocations.
We study the use of two broad classes of API invocations defined in prior work~\cite{ami2022crypto}, {\sf (1)} {\em restrictive}, which only takes a certain predefined set of parameters, and {\sf (2)} {\em flexible}, which can be implemented in an unrestricted number of ways.
We study a large number of invocations of one highly relevant and widely-studied API from each category, namely \ciphergetinstance for restrictive, and \checkServerTrusted for flexible, enabling tractable but in-depth characterization of usage. 
We also experimentally demonstrate generalizability to other APIs (Section~\ref{sec:threats}).
For an effective study grounded in the study of a rich set of crypto-API invocations, we target a security-critical app population that also makes heavy use of crypto-APIs.
That is, we study crypto-API invocations in Android apps, and more precisely, a relevant sub-population, {\em mobile-IoT apps}, \ie Android apps that serve as companion apps for IoT devices, or represent third-party integrations and automation services.
This choice enables an impactful study, as {\sf (1)} this population of apps is relatively likely to require crypto-APIs to protect sensitive IoT data at rest or in transit, {\sf (2)} the attack surface mobile-IoT apps expose as the user interfaces (UIs) and controllers of physical IoT systems is critical, and {\sf (3)} because any findings in these apps are directly relevant to a large body of prior work on crypto-API misuse~\cite{mobsf,mallodroid,rahaman2019cryptoguard,kruger2017cognicrypt,kruger2020cognicryptgen,crylogger,greenwood2014smv} that has focused on the Java Cryptographic Architecture (JCA). 
Thus, this study is initialized with $140,431$ crypto-API invocations ($118,749$ restrictive and $21,682$ flexible) extracted from $20,508$ apps from the mobile-IoT app dataset by Jin et al.~\cite{jmk+22}, and makes the following contributions:

\begin{itemize}
\item {\bf \em The study.} This is the first large scale qualitative study (n~=~$5,704$ API invocations) that characterizes the unnatural use of crypto-API in the wild. 
\item {\bf \em Complexity-based stratification for qualitative analysis.} 
We develop a {\em complexity metric} that allows us to obtain a representative sample from the broad population of $140,431$ API invocations, using complexity as a proxy for their perceived naturalness.   
This approach allowed us to obtain a sample of $5,704$ invocations ($3,599$ restrictive and $2,105$ flexible) representative of  all complexity strata. 
\item {\bf \em Qualitative Analysis and Taxonomies.} We qualitatively analyzed the $5,704$ API invocations, leveraging systematic manual reverse engineering, along with the use of tools such as Ghidra~\cite{ghidra} for investigating calls into native code, and minimal working examples to explore parameter encryption and encoding.
This extensive analysis led to the identification of $96$ unique ways in which developers express and misuse crypto-APIs, organized into {\em two taxonomies} for the restrictive and flexible category, respectively. 
\item {\bf \em Results and $17$ key findings (\fnumber{1}--\fnumber{17}).} 
We find that even seemingly conventional API invocations may not always be simple/natural, but instead involve the use of object identifiers (OIDs) and key wrapping algorithms (\fnumber{2}), encryption (\fnumber{8}), and unsafe hard-coded defaults (\fnumber{6}), and ineffective measures instead of security checks (\fnumber{12}--\fnumber{14}). 
We find several cases of {\bf \em developers evading detection tools and manual code reviews}, \eg appending arguments or manipulating strings in non-intuitive ways (\fnumber{7}, \fnumber{10}), hiding vulnerable parameters in native code (\fnumber{3}), and presenting best-practice parameters in code (\eg\ \code{AES} in \code{GCM} mode) while actually using vulnerable parameters (\eg\ \code{ECB})~(\fnumber{8}).
Finally, we show that popular crypto-API misuse detectors cannot reason about even the most basic cases from our taxonomies (\fnumber{17}), \ie of the 23 basic cases representing $59$k invocations tested against, no tool can reason about 12/23 (representing $12.8$k invocations).

\end{itemize}

To summarize, this paper characterizes the existence, diversity, and prevalence of unconventional, real-world, crypto-API invocations, and demonstrates the ineffectiveness of existing tools in reasoning about them. In doing so, it motivates the need to re-think tool design with a focus on detecting such prevalent and hard-to-find cases. 
With this objective, we distill our $17$ findings into discussion themes that lead to {\em four key takeaways}, informing future work on the challenges and opportunities in detecting unnatural crypto-API misuse, enabled by the lessons and artifacts from this work.

\section{Methodology}
\label{sec:methodology}
Our study is focused on two broad categories of crypto-API identified by prior work~\cite{ami2022crypto}: {\sf (1)} {\em restrictive} invocations, \ie where the acceptable parameter values can only fall within a predefined set of constants, and {\sf (2)} {\em flexible} invocations, where the APIs are heavily customizable, and can be implemented in any unrestricted number of ways.
For our analysis, we choose two APIs that are the canonical representations of the two key usage categories: the \ciphergetinstance API from the restrictive category, and the \checkServerTrusted method in the \trustManager interface for the flexible category.
These APIs are some of the most represented in literature, and prior work has extensively studied~\cite{ami2022crypto,jmk+22,pinningpradeep,hazhirpasand2020,nadi2016,swk+2022cambench,fahl2012eve,chenndss2024} and  designed tools to detect their misuse~\cite{egele2013empirical,fahl2012eve,kruger2017cognicrypt,kjk+2021cognicrypt,mobsf,muslukhov2018,crylogger,rahaman2019cryptoguard,zcd+2019cryptorex}.
Thus, focusing on these two relevant APIs allows for a tractable but impactful analysis.

\subsection{Extracting invocations, Generating ASTs}
\label{sec:collection}
As discussed previously in Section~\ref{sec:intro}, we target mobile-IoT apps due to their critical attack surface, and their potential of using crypto-APIs to protect sensitive IoT data. 
We were able to successfully download the latest versions of $20,508$ mobile-IoT apps from the list of ~37k previously compiled by Jin et al.~\cite{jmk+22}, as the rest were unavailable. 

We decompile the apps using jadx~\cite{jadx},{\footnote{
We experimentally confirmed that decompiled code from jadx has high fidelity with respect to the source (see the online appendix~\cite{onlineappendix}). To summarize, we compiled minimal examples of both natural and unnatural invocations observed in Sections~\ref{sec:restrictive-analysis} and~\ref{sec:flexibleresults}, using R8 and Proguard obfuscation. Decompilation resulted in code with the same characteristics as the source.}  
search the decompiled code for the signatures of the two chosen APIs using regular expressions, and extract the class files with relevant instances.
We then use the tree-sitter library~\cite{tree-sitter} (particularly its {\em S-expressions}) to generate the Abstract Syntax Tree (AST) for the entire class. 
Since tree-sitter generates concrete syntax trees that include tokens such as {\em if} in if statements, {\em commas}, and {\em parentheses} as nodes, we traverse the tree to find and keep {\em named nodes} while removing anonymous nodes. 

We traverse and filter the AST generated for the class using the relevant method signature of the relevant crypto-API.
We then extract the selected nodes with the corresponding children, and in effect, obtain the AST for each crypto-API invocation.
This approach led to a dataset of $140,431$k ASTs representing crypto-API invocations in the $20$k apps.

\subsection{Extracting a Representative Sample of Invocations using a Complexity Metric} \label{sec:comscore}
For a feasible qualitative analysis, we need a representative sample from the large dataset of all $140$k invocations of crypto-API invocations.
Given that our goal is to characterize unnaturalness in crypto-API invocation, we must obtain a sample in which all the unique populations/strata of natural or unnatural invocations are well represented.
That is, we need a way to stratify the dataset to precisely group similarly natural/unnatural invocations and sample representative examples from each group. 
This requirement motivates our metric. 

We define {\bf \em ``natural''} as the typical/conventional invocation of crypto-API according to the documentation and what tool designers expect, \eg something as {\em simple} as \code{Cipher.getInstance(\dbq{}DES\dbq{})}.
Conversely, {\bf \em ``unnatural''} indicates deviation from conventional use, which, in extremes, can be similar to the {\em complex} code in Listing~\ref{lst:desreplace}.
Hence, it is intuitive to rely on the complexity of the invocation, measured by argument size, for stratifying the dataset along the spectrum of unnaturalness.
This does not mean that more complex invocations are misuse, but instead, that {\em invocations of a similar complexity may be similarly (un)natural, and grouping them allows effectively sampling, study, and characterization}.

\myparagraph{Construction of the complexity metric}  We use the {\em S-expression} to query the AST of each crypto-API invocation and return all arguments in the argument list as seen in the source code. 
This allows us to identify the use of strings, identifiers, method calls, the ternary operator, string methods, \etc, in the API invocation.

For simplicity, we design the complexity metric based on number of arguments observed in the children of restrictive invocation nodes, \ie cipher object instantiation. 
Similarly, for flexible invocations, we use the size of the method body based on number of element in the block. 
Equation~\ref{eq:1} shows how we calculate complexity score of a method $i$, $S_i$, where $d_i$ represents the number of arguments in the method:

\begin{equation}
	S_i = \tanh[\log(d_i)]\label{eq:1} 
\end{equation}
The computation of the metric in Equation~\ref{eq:1} can be explained as follows:
Since the number of the arguments in a method, \ie $d_i$, can not be predetermined, we use the logarithmic function to normalize the argument size to reduce the disparity between the final scores. 
The second step is to use the hyperbolic tangent function, \ie $\tanh$ to calculate each complexity score between the ranges of $-1$ for least complex use to $+1$ to most complex use, since hyperbolic tangent function has a range of $-1$ to $+1$. 
That is, the $\tanh$ function enables a gradual distribution of the complexity score as the argument size increases, resulting in precise stratification.

\subsection{Qualitative Analyses} \label{sec:postanalyses}

We sample both restrictive and flexible invocations using the metric described previously for qualitative analysis.
For each invocation being analyzed, we manually examine the invocation as well as it's surrounding code context in the application. 
In many cases, we came across method invocations that required further analysis and creating minimal working examples to test our hypotheses. 
Particularly, we frequently observed native method invocation during our qualitative analysis, which we reverse engineered using Ghidra~\cite{ghidra} to identify the cipher algorithms and other arguments in native code. 
Finally, some invocations were more challenging to understand as they involved handcrafted transformations, including encoding and/or encryption. 
To understand these cases, we used our intuition to develop hypotheses regarding the eventual effect of the transformations, and tested the hypotheses by manually constructing minimal working examples.

\subsection{Prevalence Analysis} 
Our prevalence analysis seeks to obtain a lower bound on the presence of each of the unique unnatural crypto-API invocations found using qualitative analysis  across the entire dataset. 
We do not seek to find all instances of the said usage. Instead, our approach only marks the ones that it is most confident about, giving us a conservative estimate.
We used a combination of signatures from argument types retrieved from the AST after parsing and signatures from specific method calls to measure prevalence. 
For example, the signature: \code{\{'identifier': 3, 'ternary\_expression': 1\}} indicates the use of three identifiers with the ternary operator, resulting in the identification of \code{Cipher.getInstance(z2 ? CBC\_PADDING : CBC\_NOPADDING)} and similar usage.

\section{Analysis of Restrictive Invocations}
\label{sec:restrictive-analysis}

\begin{figure}[t]
	\def\arraystretch{1.5}
	\centering
	\includegraphics[width=3.2in]{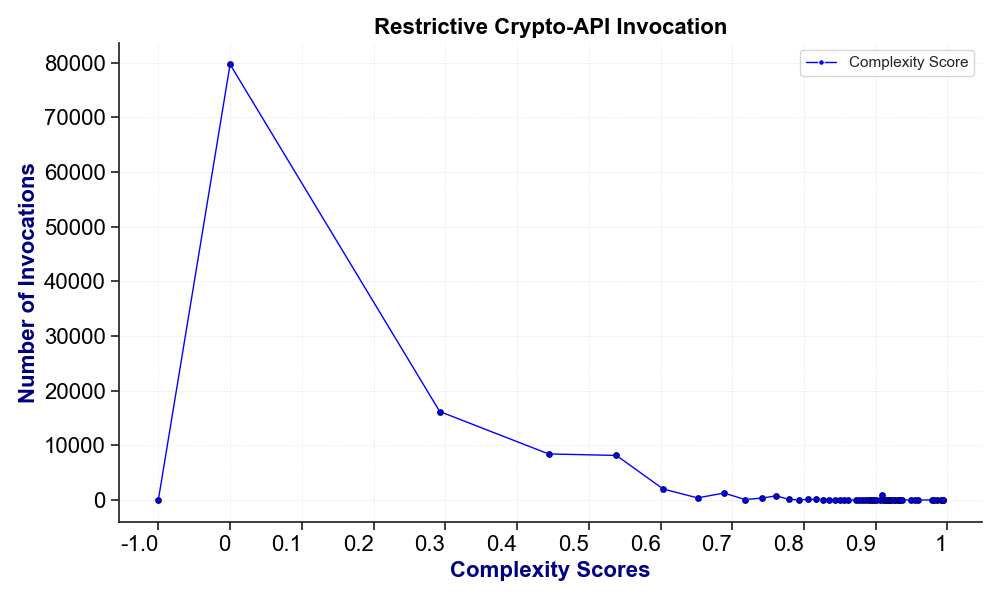}
	\caption{{\small {Restrictive Invocation Complexity Score Distribution }}} 
		\label{fig:restrictive}
\end{figure}

We extracted $118,749$ invocations of the restrictive API chosen for this study, \ie\ \cipherget. 
Applying the complexity metric to these invocations resulted in $55$ distinct  complexity scores.
The distribution of the invocations across the complexity scores is shown in Figure~\ref{fig:restrictive}. 

As seen in Figure~\ref{fig:restrictive}, $79,671$ (67\%) attain a score of ``0'' (\ie henceforth noted as \score{0}).
At face value, this may be interpreted as showing that developers mostly use \cipherget in the most conventional way by specifying a single string, \eg \texttt{\small Cipher.getInstance(\dbq{}AES/GCM/NoPadding\dbq{})}.
However, our qualitative analysis (Section~\ref{sec:restrictive-results}) demonstrates otherwise, that even seemingly natural invocations may be quite unnatural.

\subsection{Obtaining a Representative, Stratified Sample of Restrictive Invocations}
\label{sec:restrictive-sampling}
For a feasible but in-depth qualitative analysis, we use stratified random sampling to obtain a representative sample of invocations to analyze for each of the $55$ distinct complexity scores.
We treat score as an independent population, \ie a stratum. 
For each stratum, we calculate the number of samples we need to study for results representative of the entire stratum using the Qualtrics sample size calculator~\cite{qualtrics}, assuming a 5\% error margin and 95\% confidence, both standard values. 
For scores with a limited population  (\eg\ \score{-1}), we sampled the entire strata, whereas for others that were most populated, we obtained a relatively small but representative sample (\eg\ \score{0}).
Our online appendix~\cite{onlineappendix} shows the size of the representative samples for each stratum.
Using this approach, we selected a total of $3,599$ representative invocations of \cipherget for qualitative analysis. 

\begin{figure*}[t]
	\def\arraystretch{1.5}
	\centering
	\includegraphics[width=6in]{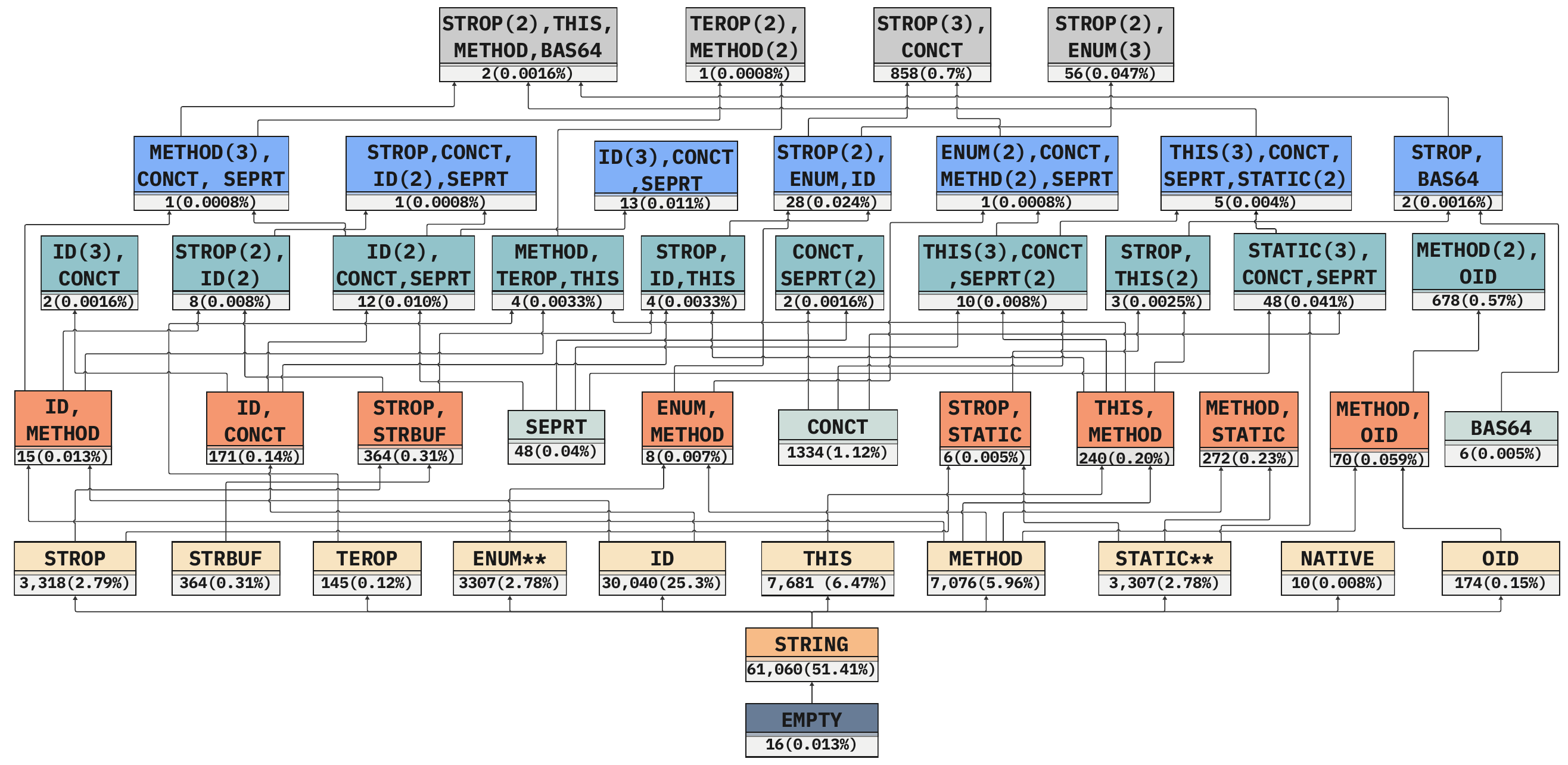}
	\caption{{\small {Taxonomy of Unnatural Restrictive Crypto-API Invocations with Prevalence information.}}}
	\label{fig:tax_res}
\end{figure*}

\subsection{Results and Findings from the Qualitative Analysis of Restrictive Invocations}
\label{sec:restrictive-results}
Our analysis of the $3,599$ representative invocations allowed us to characterize the unique ways in which developers express restrictive crypto-APIs.
We represent this characterization in a {\em taxonomy of unnatural restrictive crypto-API use}, as shown in Figure~\ref{fig:tax_res}.
The taxonomy shows $15$ {\em basic} ways in which developers express the \cipherget API, and $29$ {\em complex} ways in which they further combine the basic ways, resulting in $44$ total taxonomy labels.
Table~\ref{tab:res_tax} in the Appendix describes the 15 basic labels. 
 
The rest of this section describes the results of our qualitative analysis, wherein we discuss individual scores or score ranges with key results using salient examples. 

\noindent

\myparagraph{[SCORE = -1] -- No arguments provided}
Since the metric was designed using a $\tanh$ function which has a range of $-1$ to $+1$, 
we get a \score{-1} for \cipherget with no arguments provided. 
There were $16$ invocations of this type, expressed as {\bf \em type}{\tt \small Cipher.getInstance()}, where the {\em \textbf{type}} can take the following values: "Block", "AES", and "XML", each value resulting in a unique default cipher configuration.

To elaborate, {\bf \em Block}\code{Cipher.getInstance()} defaults to AES in GCM mode. 
Similarly, using {\bf \em AES}\code{Cipher.getInstance()} defaults to AES in CBC mode \ie\ \code{AES/CBC/PKCS5Padding}, which is vulnerable in certain situations~\cite{aescbc}. 
\finding{When instantiated without arguments, the outcome of \cipherget depends on the {\em type}, which may lead to vulnerable defaults.}
\myparagraph{ [SCORE = 0] --  String Literals, and Identifiers} 
Using string literals such as \code{``AES/GCM/NoPadding''} as the parameter is a natural and expected use of API such as \cipherget. 
Indeed, our prevalence analysis shows $60,060/118,749$ (or $51.41\%$) invocations of this kind.
However, we observe that developers may use non-intuitive literals, potentially leading to undetected misuse.

We observed several cases where developers used object identifiers (OIDs) corresponding to common ciphers, particularly the use of {\tt \small {\dbq{}1.2.840.113549.3.2\dbq{}}}, which resolves to the vulnerable RC2 cipher (see RFC~2268~\cite{rfc2268}).  
Further, we observed the use of the AES Key Wrap Algorithm without padding ({\em "AESWrap"}) and with Padding ({\em "AESKWP"}), both of which default to ECB mode~\cite{rfc3394}~(\eg in default providers for Java and Android such as SUNJCE~\cite{sunjce}), which is vulnerable. 
Our prevalence analysis establishes a lower bound of $174$ uses of OIDs, and $1,487$ of AESWrap. 

\finding{Restrictive invocations with String literals as parameters may seem natural from a syntactic standpoint but use non-intuitive abstractions such as OIDs and key wrapping algorithms with vulnerable defaults. Such semantic unnaturalness may escape detection tools.} 
After string literals, the use of identifiers is another natural way in which developers specify parameters. 
Since identifiers can be assigned to any variable and method, we perform additional analysis to characterize their use. 
We found several unique and non-trivial ways in which developers use identifiers, each of which occupy separate spots in the taxonomy. 
First, we found $4$ cases where a vulnerable block chaining mode was used, \ie\ \code{ECB} in Android API levels $< 23$, and a secure mode, \ie\ \code{GCM} for higher APIs. 
The arbitrary cut-off at API 23 is odd as \code{GCM} has been supported since API $10$~\cite{androidgcmcipher}. 

We also found invocations where the identifier holds the result of arbitrary string manipulation, \eg\ \code{Cipher.getInstance{(<identifier>)}}, where \code{<identifier>} = \code{String.format(\dbq\%s/\%s/\%s\dbq,\dbq{}AES\dbq{},\dbq{}CBC\dbq{},\dbq{}PKCS7Padding\dbq{})}, vulnerable to oracle padding attacks~\cite{aescbc}.
Our prevalence analysis found $44$ instances of this type. 
Further, developers may also assign identifiers to methods that perform complex operations such as \code{XOR} and \code{Base64} encoding/decoding on cipher arguments. 
This may be prevalent in the wild as similar code was found on stackoverflow~\cite{stackoverflowxor} as early as $2011$. 
Instances of such \code{Base64} decoding and/or \code{XOR} use can be seen in Listing~\ref{lst:base64} (also Listing~\ref{lst:xor1} and Listing~\ref{lst:cbctoecb} in the Appendix).  
Our prevalence analysis confirmed $6$ instances of such complex operations, although this is a conservative estimate as detecting such use is a challenging.
\noindent
\begin{minipage}{\linewidth}
\begin{lstlisting}[frame=tb,caption={{\small Native Method Call with Identifier 1.}}, label={lst:native1},language=java]
private static native String nativeGetString(int i);
nativeGetString = nativeGetString(1); 
Cipher.getInstance(nativeGetString);
\end{lstlisting}
\end{minipage}

\noindent	
\begin{minipage}{\linewidth}
\begin{lstlisting}[frame=tb,caption={{\small Native Method Call with Identifier 2.}}, label={lst:native2},language=java]
private final native String getCypherTransformation();
f6401b = crypto.getCypherTransformation();
Cipher.getInstance(f6401b)
\end{lstlisting}
\end{minipage}

Finally, developers frequently assigned identifiers to native method calls, which may conceal misuse.
The examples shown in Listing~\ref{lst:native1} and Listing~\ref{lst:native2} demonstrate the common patterns of native method calls to retrieve string arguments.
There is no performance-based rationale for such calls to native methods that merely retrieve string parameter values to use in a the Java crypto API, as the JNI call results in performance degradation. 
Our prevalence analysis identified $10$ such cases, all of which were misuse, including $5$ uses of \texttt{AES/ECB/NoPadding}, $1$ use of \texttt{AES} which defaults to \code{ECB} mode in JCA, and $4$ uses of \texttt{AES/CBC/PKCS5Padding}.  

\finding{While identifier use in place of a string literal in restrictive crypto API may seem natural, we observe several types of non-intuitive usage, such as string manipulation or complex computation, most of which lead to misuse.
Importantly, we observe JNI calls to obtain vulnerable parameters from native code to be eventually used in Java API, showing signs of  
{\em evasive} API misuse.}
\myparagraph{[SCORE = 0.4439] -- Method calls, enum types, constructors}
This score is characterized by the use of enum type, constructors, method calls and "this" keyword to reference the current object in a method. 
Note that the method calls here are classified distinctly as they were found in \cipherget invocations, unlike the prior discussion in \score{0} of method calls used to initialize identifiers used in the invocations. 
Out of the $368$ invocations analyzed, we were able to identify the precise cipher used in $21$ instances as we stopped after two levels of indirection, or if the value was retrieved from the network. $
20/21$ of these parameters were vulnerable (\eg using \DES). 
Our prevalence analysis showed that referencing the current object using the "\code{this}" keyword accounts for $6.47\%$ or $7,681$ instances.

\finding{The use of enum types, method calls, and constructors leads to complex call chains that are hard to resolve due to severe indirection and network-derived values. 20/21 of the parameters we could resolve were vulnerable, highlighting the need to analyze such cases.}
Among the method calls we analyzed, one in particular led to a ternary operator that used an arbitrary integer value to switch between the \code{AES} in \code{GCM} mode, and the default (which results in \code{ECB} mode), as shown in Listing~\ref{lst:ternary1} below. 
\noindent
\begin{minipage}{\linewidth}
	\begin{lstlisting}[frame=tb,caption={{\small Use of Ternary Operator via Method Call.}}, label={lst:ternary1},language=java]
Cipher.getInstance(getTransformation()) 
-----------------------------------------
private static String getTransformation() {
 return f882fB >= 2 ? "AES/GCM/NoPadding" : "AES";
	\end{lstlisting}
\end{minipage}
\finding{Developers may use enum type constant and ternary operator to store cipher algorithm and switch between them based on arbitrary conditions such as constant values. However, the switch between secure and vulnerable parameters indicates that developers recognize the difference, but misuse on purpose.}
\myparagraph{[SCORE = 0.5385] -- Ternary Operator, String Buffer, and Native Method Invocations}
While we have seen the {\em indirect} use of ternary operator via method calls in previous score, this score is characterized by the {\em direct} use of the ternary operator, native method calls, and string buffers as arguments for cipher object instantiation.

\noindent
\begin{minipage}{\linewidth}
	\begin{lstlisting}[frame=tb,caption={{\small Direct Native Method Call.}}, label={lst:nativedirect},language=java]
Cipher.getInstance(requestTransform(1));
public static native String requestTransform(int i);
		\end{lstlisting}
	\end{minipage}
To elaborate, we found direct calls from \cipherget instantiations to native methods that retrieve cipher transformation strings and values. 
For example, the native method \textit{requestTransform} from Listing~\ref{lst:nativedirect} is defined as follows: if the value of the parameter $i\neq0$, it returns \code{AES/ECB/NoPadding}, otherwise, it returns \code{AES/CBC/PKCS5Padding}. 
However, as seen in Listing~\ref{lst:nativedirect}, $i$ is hardcoded to $1$, which defaults to \code{ECB}. 
We found $5$ such cases of misuse in $5$ different apps. 

\finding{Developers use conditions based on integers to switch between a vulnerable and good parameters. However, the integer value may be hardcoded, making the vulnerable parameter the default.}
Further, we found several direct uses of the ternary operator in a cipher object instantiation. 
Particularly, we observed instances where the operator was dependent on arbitrary integer values (often from the network), as seen before in Listing~\ref{lst:ternary1}.

	\noindent
\begin{minipage}{\linewidth}
	\begin{lstlisting}[frame=tb,caption={{\small Use of StringBuffer.}}, label={lst:stringbuffer1},language=java]
StringBuffer stringBuffer = new StringBuffer();
stringBuffer.append("DESede/CBC/");
stringBuffer.append("NoPadding");
this.f13712c = Cipher.getInstance(stringBuffer.toString());
	\end{lstlisting}
\end{minipage}
Finally, we find the string buffer being used to append fragments of the cipher algorithm. 
For example Listing~\ref{lst:stringbuffer1} shows a code fragment where the developers split the cipher algorithm and cipher mode separate from the padding mode. 
This is a relatively simple instance of this type of invocation expressed in a string buffer, as the next score shows more complex cases that involve non-intuitive splitting and combining of values. 
However, such simple use of string buffers may prove challenging for tools to detect as the manner in which developers may split the algorithm may be non-trivial. 
Our prevalence analysis found $364$ such instances. 

\myparagraph{[SCORE = 0.6 -- 0.69] -- Disguising Vulnerable Cipher choices via String Builders, Encoding, Encryption}
The general structure of our observed invocations in this score range is similar to prior scores involving ternary operators, method calls, and the use of string builders/buffers. However, their use and parametrization is quite different as the parameter string is irregularly fragmented.
	\noindent
\begin{minipage}{\linewidth}
	\begin{lstlisting}[frame=tb,caption={{\small Use of String Builder.}}, label={lst:stringbuilder},language=java]
StringBuilder sb = new StringBuilder();
sb.append("AES");
sb.append("/EC");
sb.append("B/PKCS7P");
sb.append("adding");
Provider provider = Security.getProvider("BC");
cipher = Cipher.getInstance(sb.toString(), provider);
	\end{lstlisting}
\end{minipage}
Consider the example in Listing~\ref{lst:stringbuilder}, which demonstrates the fragmentation of \code{AES/ECB/PKCS7Padding} into odd fragments such as \code{/EC} and \code{B/PKCS7P}.
This type of irregular fragmentation may not only present a challenge to misuse detectors, but may also present additional evidence of developers trying to evade tools and/or code reviews by expressing their choice of vulnerable parameters in non-intuitive ways.
Our prevalence analysis found $364$ such uses. 
\finding{String operations techniques such as use of \code{StringBuffer} and \code{StringBuilder} may be used to append irregular fragments of algorithm name, mode and padding. Developers may use this unconventional design to {\em evade} detection tools.}
We found further complex uses that leveraged encryption, encoding, and XORs to conceal vulnerable algorithm values (\eg transforming \code{CBC} to \code{ECB} at runtime using XOR, see Listing~\ref{lst:cbctoecb} in the Appendix). 
A salient example is in Listing~\ref{lst:decode} below, where the secure algorithm appears in code (\ie\ \code{AesGCMUtils}), but a hidden (encrypted), vulnerable, value is eventually used:  
\noindent
\begin{minipage}{\linewidth}
	\begin{lstlisting}[frame=tb,caption={{\small Using Secure Cipher to Encrypt Insecure Cipher.}}, label={lst:decode},language=java]
algo_encode = "32Bi2A5oaH61xilScou92x9faAiO0SOBXmb0X/wqAijapt8K"; 
Cipher.getInstance(AesGcmUtils.decode(algo_encode));
	\end{lstlisting}
\end{minipage}
In Listing~\ref{lst:decode}, the developer uses Base64 to decode a string, which is then decrypted using \code{AES/GCM/NoPadding} (\ie\ \code{AesGcmUtils}) and a hardcoded key. 
However, this decryption results in \code{AES/ECB/NoPadding}, which eventually serves as the vulnerable parameter (see full flow in Listing~\ref{lst:fulldecrypt} in the Appendix).
To automated or and manual security analysis, this usage will appear to be safe, since GCM mode is visible in code, and ECB is not, when in fact it is vulnerable.
Our prevalence analysis identified $15$ instances of such evasive misuse.
\finding{Developers use encryption, encoding, and XORs to conceal vulnerable parameter choices.
In particular, developers may {\em use secure parameters to conceal the vulnerable parameters} in an attempt to misdirect automated and manual code reviews.}

\myparagraph{[SCORE = 0.7 -- 0.79] -- Even more complex instances of previously identified usage}
In this score range, the cryptographic-API {\em (mis)use} we found were similar to what we saw previously; however, they were more complex, particularly in terms of containing additional arguments, which led to a higher complexity score. 

	\noindent
\begin{minipage}{\linewidth}
	\begin{lstlisting}[frame=tb,caption={{\small Directly Decoding the Cipher Algorithm.}}, label={lst:base64},language=java]
Cipher.getInstance(new String (Base64.decode("REVTL0NCQy9QS0NTNVBhZGRpb mc=",2)));
	\end{lstlisting}
\end{minipage}

For example, we found the use of \code{Base64} for decoding a string that was then passed as a direct argument to \cipherget as shown in Listing~\ref{lst:base64} (6 instances of this type), unlike the previous findings, which were either through method calls or identifiers. 
We observed similar cases including the use of ternary operators ($62$ instances), string operation techniques such as \code{format} that may be combined with use of identifiers and other string operations ($8$ instances), a more fragmented concatenation, \ie concatenating individual elements including ``/'' ($65$ instances), and more complex method chaining. 

\finding{For restrictive invocations, crypto-API usage found indirectly used in lower complexity scores, \ie as identifiers and method invocations, were observed directly used in invocations with higher complexity.}
\myparagraph{[SCORE = 0.8 -- 0.89] -- Nested Ternary Operator \& Separator}	
We found a use of nested ternary operators as direct arguments, where one ternary operator checks if the \code{Build.Version.SDK\_INT} is lower than Marshmallow, after which a string that signifies either case is selected, as seen in Listing~\ref{lst:slash} in the Appendix.
If the resultant string is "M", the next operator selects \code{RSA/ECB/OAEPWithSHA-1AndMGF1Padding}, whereas if it is "PREM" it selects \code{RSA/ECB/PKCS1Padding}.  

\myparagraph{[SCORE~=~0.9~--~1.0] -- Highly complex usage and obfuscation}
On the lower end of this score range i.e. {\em \underline{SCORE \#0.908438}}, we find a highly complex crypto-API misuse that uses string manipulation. 
As shown in Listing~\ref{lst:charat}, the \code{charAt} method is used to select specific characters, along with the addition operator $"+"$ to add the selected characters, which eventually results in \code{\dbq{}AES/ECB/NoPadding\dbq{}}. 
However, the only value shown in the code is \code{AES/GCM/NoPadding}, \ie this particular instance is yet another example of {\em evading detection tools and manual code reviews}, wherein the developer effectively uses \code{ECB} mode, but makes it appear as they they use \code{GCM} mode. 
Our prevalence analysis identified $858$ instances of this pattern, which was also seen in prior work~\cite{mao+24}. 
\noindent
\begin{minipage}{\linewidth}
\begin{lstlisting}[frame=tb,caption={{\small Using charAt to Manipulate String.}}, label={lst:charat},language=java]
Cipher.getInstance("AES/" + ((char) ("AES/GCM/NoPadding".charAt(4) - 2)) + "AES/GCM/NoPadding".charAt(5) + ((char) ("AES/GCM/NoPadding".charAt(6) - 11)) + "/NoPadding")
\end{lstlisting}
\end{minipage}
\finding{Developers may evade detectors and manual reviews using complex string manipulation that presents a benign value, but resolves to a vulnerable one.}
On the higher end of this score range, we begin to find signs of obfuscation and parsing errors. 
Unicode characters in Java can be distinctively identified by use of a leading"$\backslash$u" and in addition to complex mathematical operations in the arguments, we assume this type of code is heavily obfuscated.

\section{Analysis of the Flexible Invocations}
\label{sec:flexibleresults}

We observed $21,682$ invocations of the \checkServerTrusted API in the 20k mobile-IoT apps. 
Figure~\ref{fig:flexible} shows their distribution across the $339$ complexity scores.
The \score{-1} corresponds to an empty \checkServerTrusted method body, and was attained by $34.2\%$ ($7,417$/$21,682$) invocations.
Further, about 20\% of the invocations contribute to $309$ distinct scores toward the higher end of complexity (\ie \score{0.9} and above).
This indicates that while the majority of the invocations are rather simple,  developers may implement such flexible API in several unique and highly complex ways.
\begin{figure}[t]
	\centering
	\includegraphics[width=3.2in]{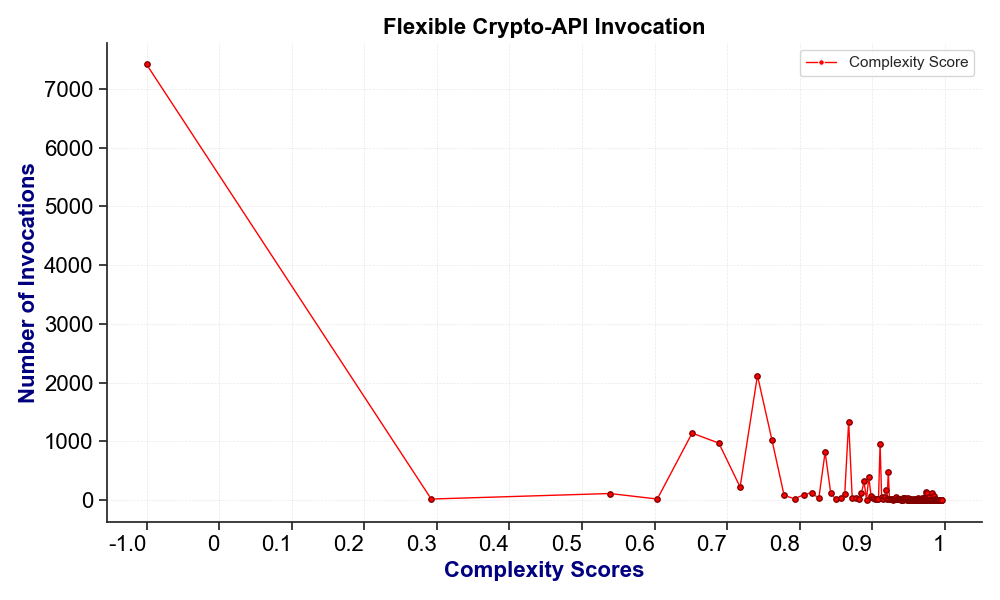}
	\caption{{\small {Flexible Invocation Complexity Score Distribution.}}} 
	\label{fig:flexible}
\end{figure}

\subsection{Obtaining a Representative, Stratified Sample of Flexible Invocations}

As there are $339$ distinct complexity scores for flexible invocations, many due to only one invocation, considering each distinct score as a stratum is infeasible.
Therefore, for a viable qualitative analysis, we treated each score {\em range} as an individual stratum, starting the first range at \score{0} and considering ranges at intervals of $0.1$ (\ie $0$ -- $0.09$, $0.1$ -- $0.19$, ..., $0.9$ -- $1$). 
We consider the special case with \score{-1} (\ie empty method body) as a unique strata. 

We calculate the number of representative samples for each score range (\ie stratum) using the Qualtrics sample size calculator~\cite{qualtrics}, assuming a margin of error of 5\%, and confidence of 95\%. 
Our online appendix~\cite{onlineappendix} provides details on each stratum, its population, and the size of the representative sample for the stratum/score range. 
Using this approach, we selected a total of $2,105$ representative invocations of \checkServerTrusted for qualitative analysis.

\begin{figure*}[t]
	\def\arraystretch{1.5}
	\centering
	\includegraphics[width=7in]{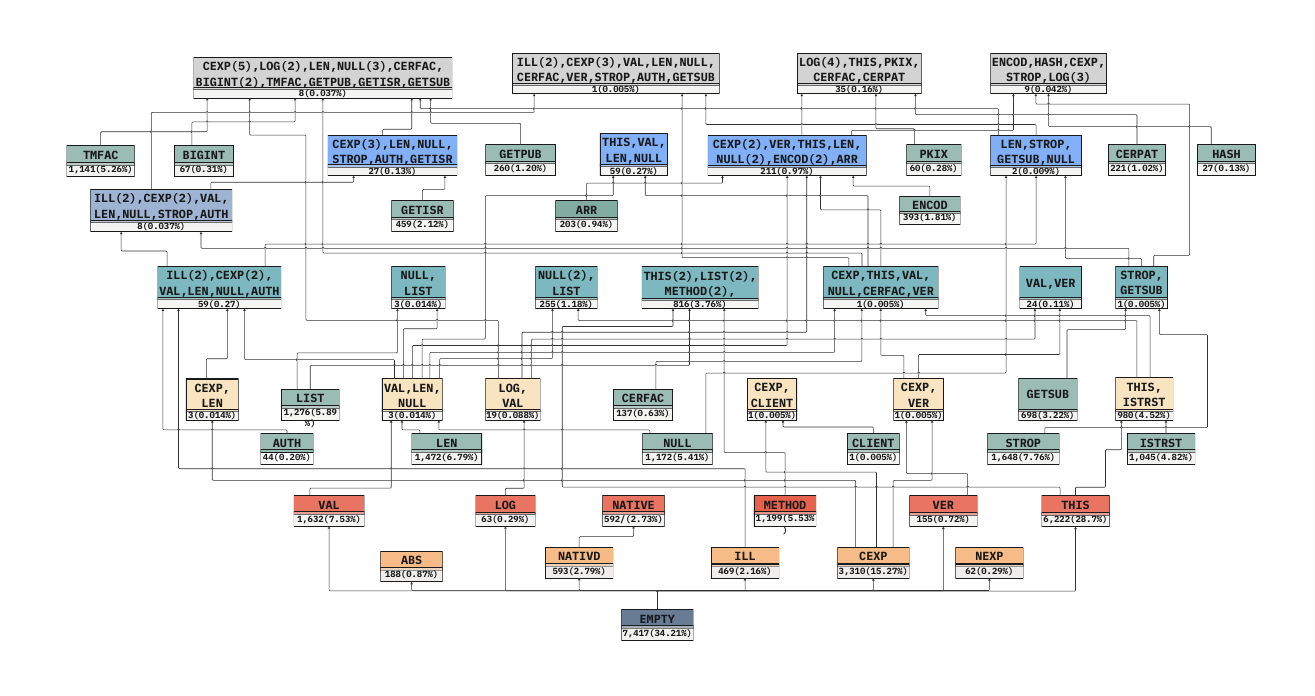}
	\caption{{\small {Taxonomy of Unnatural Flexible Crypto-API Invocations with Prevalence information.}}}
	\label{fig:tax-flexible}
\end{figure*}

\subsection{Results and Findings from the Qualitative Analysis of Flexible Invocations}
Our analysis of the $2,105$ representative invocations helped us characterize the unique ways in which developers express flexible crypto-APIs, represented in the form of a {\em taxonomy of unnatural flexible crypto-API use}, as shown in Figure~\ref{fig:tax-flexible}.
We identified $30$ {\em basic} ways in which developers implement certificate validation logic via \checkServerTrusted API, and $22$ {\em complex} ways in which they further combine the basic ways, resulting in $52$ total taxonomy labels.
Table~\ref{tab:flex_tax} in the appendix elaborates on the 30 basic labels. 

The rest of this section describes the results and findings from our qualitative analysis of flexible API invocations. 
We discuss score ranges with key results, providing salient examples along the way.

\myparagraph{[SCORE = -1] -- Empty Method Body}
All implementations with this score have an empty method body, which causes the \trustManager to accept all certificates. 
Our prevalence analysis shows that this accounts for $7,417$ instances.

\finding{Vulnerable \trustManager implementations with empty \checkServerTrusted methods are highly prevalent, echoing findings from prior work~\cite{fahl2012eve,rahaman2019cryptoguard}}

\myparagraph{[SCORE = 0.6 -- 0.7] -- Calling another method}
This score range is characterized by a method call from \checkServerTrusted, such that the called method is responsible for certificate validation. 

\noindent
\begin{minipage}{\linewidth}
	\begin{lstlisting}[frame=tb,caption={{\small Native Method Call from Java Method.}}, label={lst:nativemeth},language=java]
public void checkServerTrusted(X509Certificate[] x509CertificateArr, String str) {
 n_checkServerTrusted(x509CertificateArr, str); }
	\end{lstlisting}
\end{minipage}
Besides calls to other methods in Java, we also found calls to  native methods, such as \code{n\_checkservertrusted} in Listing~\ref{lst:nativemeth}. 
Our prevalence analysis reveals that native method calls account for $592$ instances and other/Java method calls account for $1199$ instances. 
As such invocations rely on other called methods for the certificate validation,  the effectiveness of such method may not be apparent to security tools unless they analyze the called methods. 

\finding{Only analyzing checks in the target method (\eg\ \checkServerTrusted) is inadequate when analyzing crypto-API misuse in {\em flexible} invocations, as they rely on other native/Java methods.}

\myparagraph{[SCORE = 0.7 -- 0.8] -- Certification Expiration}
This score range is characterized by calls to the \code{checkValidity} method to check the certificate for expiration.
Listing~\ref{lst:val} shows a typical implementation of this type of invocation. 
	\noindent
\begin{minipage}{\linewidth}
	\begin{lstlisting}[frame=tb,caption={{\small Checking expiration only.}}, label={lst:val},language=java]
public void checkServerTrusted(X509Certificate[] x509CertificateArr, String str) throws CertificateException {
	for (X509Certificate x509Certificate : x509CertificateArr) {
		x509Certificate.checkValidity();}}
	\end{lstlisting}
\end{minipage}
It is important to note that this type of invocation accepts any unexpired certificate, and is not a substitute for certificate verification.
Our prevalence analysis shows that out of the $1,632$ instances of the use of the \code{checkValidity} method in \checkServerTrusted, $62.5\%$ have contain no other checks, and hence, accept all unexpired certificates.

\finding{The check for expired certificates, if present, is commonly the {\em only check} in such invocations, resulting in {\em any} unexpired certificates getting accepted.}

\myparagraph{[SCORE = 0.8 -- 0.9] -- TrustStrategy, Ineffectual Implementations}
We found \code{TrustStrategy.isTrusted} method from Apache {\em httpclient} being used for verifying certificates, often in the absence of any other certificate verification logic. 
Such use is vulnerable, as the \code{TrustStrategy.isTrusted} method does not consult the \trustManager that is configured in the \code{SSLContext}~\cite{istrusted}, which may void the SSL pinning configured by the developer~\cite{codeistrusted}.
Our prevalence analysis found $1,045$ such instances. 
\noindent
\begin{minipage}{\linewidth}
	\begin{lstlisting}[frame=tb,caption={{\small Complex implementation without checks.}}, label={lst:client},language=java]
public void checkServerTrusted(X509Certificate[] x509CertificateArr, String str) throws CertificateException {
 try {checkClientTrusted(x509CertificateArr, str);
 } catch (Exception e10) {
  throw new CertificateException("Certificate not trusted. It has expired", e10);}
 ...
 public void checkClientTrusted(X509Certificate[] x509CertificateArr, String str) throws CertificateException {}
	\end{lstlisting}
\end{minipage}	
In this score range, we also observed instances that had relatively complex implementations {\em that did not do anything}.
Listing~\ref{lst:client} shows a typical example, where \checkServerTrusted calls the \code{checkClientTrusted} method, however, \code{checkclienttrusted} is empty, which  in turn allows any certificate to be accepted. 
	\noindent
\begin{minipage}{\linewidth}
	\begin{lstlisting}[frame=tb,caption={{\small Logging certificate only.}}, label={lst:log},language=java]
public void checkServerTrusted(X509Certificate[] x509CertificateArr, String str) {
 for (X509Certificate x509Certificate : x509CertificateArr) {
  Log.e(TcpClient.TAG, "Certificate:" + x509Certificate);}}
	\end{lstlisting}
\end{minipage}
While specific implementations may vary from the one in Listing~\ref{lst:client}, we found several such ineffectual implementations that seem non-empty and complex, but do not conduct any relevant checks on the certificate chain.
For example, Listing~\ref{lst:log} shows a method that simply logs all the certificates in the certificate chain. 
Our prevalence analysis found $63$ instances of such invocation.
\finding{The presence of complex code in \checkServerTrusted may not translate to effective certificate checks. Instead, developers may simply log the received certificate, or simply call one or more other method that is empty.}
\myparagraph{[SCORE = 0.9 -- 1.0] -- String Operations}
One of the most prominent observations in samples falling in this score range is the use of string comparison/manipulation methods such as 
\code{\dbq{}.equals\dbq{}}, \code{\dbq{}Arrays.equals\dbq{}}, \code{\dbq{}list.add\dbq{}} and \code{\dbq{}.contains\dbq{}}.
These methods were commonly found to be used to compare values from the certificate chain, such as the encoded form of a root certificate or a public key, and the distinguished names of the issuer or subject, with values hard-coded in the application.
Listing~\ref{lst:materials} lists the common materials used for such comparisons, whereas Listing~\ref{lst:extract} shows how these values are compared. 
\noindent
\begin{minipage}{\linewidth}
\begin{lstlisting}[frame=tb,caption={{\small  Certificate Materials used for Validation.}}, label={lst:materials},language=java]
x509CertificateArr[0].getencoded()
getIssuerDN()
getIssuerDN().getName()
getSubjectDN().getName()
x509Certificate.getPublicKey().getEncoded()
\end{lstlisting}
\end{minipage}

		\noindent
\begin{minipage}{\linewidth}
	\begin{lstlisting}[frame=tb,caption={{\small Using ".contains" to validate certificates.}}, label={lst:extract},language=java]
x509Certificate.getSubjectDN().toString().contains
("EMAIL ADDRESS=sales-usa@extron.com, CN=Quantum Ultra, OU=Engineering, O=ExtronElectronics, L=Anaheim, ST=CA, C=US")
	\end{lstlisting}
\end{minipage}		 
Our prevalence analysis identified $1,648$ instances of the use of string operation methods described above. 
In addition, we also observed the use of list interfaces ($1,276$ instances) and array methods ($203$ instances) to manage certificate materials.
Thus, we can draw a parallel between restrictive and flexible invocations, as string operations seems to be a common theme in both categories of crypto-API invocations. 

\finding{Both restrictive and flexible invocations of crypto-API have the use of string operation methods, method calls, and native code in common.}

Finally, our prevalence analysis found $459$ instances of IssuerDN and $698$ instances of SubjectDN used in such comparisons, even though both are denigrated~\cite{androidx509cert, javax509cert}. 
Developers also use \code{SHA1}, which is vulnerable to collisions, to compare values of certificate materials (\eg an encoded public key or root certificate) with hard-coded values {\em to establish trust or pinning}. 
An example of this is shown in Listing~\ref{lst:sha1getenc} in the Appendix.
Our prevalence analysis finds $43$ instances of \code{SHA1} use in similar certificate validation contexts, $21$ instances of \code{SHA1} use by itself (\ie not in such contexts), and $6$ of \texttt{SHA256} being used in such certificate validation contexts. 
Additionally, developers also hard code specific certificate chain material for validation purposes, as shown  Listing~\ref{lst:extract}. 

\finding{Developers use denegrated methods such as \code{getSubjectDN} and \code{getIssuerDN}, and deprecated algorithms (\code{SHA1}) in certificate validation.}

\section{The Effectiveness of Tools at Detecting Misuse in Real-world Crypto-API Invocations}
\label{sec:discussion-tools}
The observations from our analysis regarding real-world crypto-API usage motivate a critical question for vulnerability detection in practice: {\em how effective are popular detection tools at finding misuse in real-world invocations, particularly those exhibiting the unnaturalness characterized in this study?}
 
\myparagraph{Experimental Setup} To answer the this question, we isolated a set of 23 of the most {\em basic} cases from a total 96 in our taxonomies, such that if a tool fails to detect these, it will fail to detect any of the other more unnatural cases. 
These 23 cases represent a total of $59,002$ instances of observed crypto-API invocations, \ie $45,789$ restrictive and $13,213$ flexible invocations.
We developed minimal APKs with these 23 cases implemented using invocations with vulnerable parameters found in our study. 
For some cases, such as \code{ID}, we implemented more than one vulnerable invocation as there were multiple possible variants from our dataset. 
We then tested 6 popular tools: CryptoGuard~\cite{cryptoguardtool}, CogniCrypt~\cite{cognicryptsast}, SpotBugs~\cite{spotbugstool,findsecbugs}, CodeQL~\cite{codeql}, Semgrep~\cite{semgrep}, and MobSF~\cite{mobsf}.

\begin{table}[t!]
\scriptsize
\centering
\def\arraystretch{1.5}
\caption{\small Detection of 23 basic cases from the taxonomies by popular crypto-API misuse detection tools.}
\setlength{\tabcolsep}{0.5pt}
\begin{tabularx}{\columnwidth}{p{0.12in}|l|c|c|c|c|c|c}
\Xhline{2\arrayrulewidth}
\multicolumn{7}{l} {\textbf{Restrictive Invocations}}	\\
\hline
 & \textbf{\em Label}	&	\textbf{CCRYPT}	&	\textbf{CGUARD}	&	\textbf{SPOTBUGS}	&	\textbf{CODEQL}	&	\textbf{SEMGREP}	&	\textbf{MOBSF}\\ \hline
1. & \textbf{STRING/OID}&	\greencheckmark	&		\redcheckmark	&	\greencheckmark	&	\greencheckmark	&		\redcheckmark	&		\redcheckmark \\ 
\hline
2. & \textbf{ID}&	\greencheckmark	& \greencheckmark &	\greencheckmark	&	\greencheckmark	&	\ding{119}	&	\redcheckmark \\ 
\hline
3. & \textbf{METHOD}&		\graycheckmark 	& 	\redcheckmark &		\redcheckmark	&		\redcheckmark	&		\redcheckmark	&		\redcheckmark \\ 
\hline
4. &  \textbf{METHOD*}	& 	\graycheckmark &		\redcheckmark	&		\redcheckmark	&		\redcheckmark	&		\redcheckmark	&		\redcheckmark	\\ 
\hline
5. &\textbf{NATIVE}&		\graycheckmark	&		\redcheckmark&		\redcheckmark	&		\redcheckmark	&		\redcheckmark	&	\redcheckmark	\\ 
\hline
6. & \textbf{STROP} &		\graycheckmark	&		\redcheckmark	&		\redcheckmark&	\ding{119}	&		\redcheckmark	& 	\color{red} \ding{54}  \\ 
\hline
7. & \textbf{STRBUF}&		\graycheckmark	&\ding{119}		&		\redcheckmark	&		\redcheckmark	&		\redcheckmark&		\redcheckmark	 \\ 
\hline
8. & \textbf{STRBL*}	&		\graycheckmark	&	\ding{119}	&		\redcheckmark	&		\redcheckmark	&		\redcheckmark	&		\redcheckmark \\
 \hline
9. &\textbf{CONCT}	&	\greencheckmark	&	\greencheckmark	&	\greencheckmark	&	\ding{119}	&	\ding{119}	&\greencheckmark \\ 
\hline
10. &\textbf{BAS64}	&		\graycheckmark	&		\redcheckmark	&		\redcheckmark	&		\redcheckmark	&		\redcheckmark	& 	\redcheckmark \\ 
\hline
11. & \textbf{ID,METHOD}&		\redcheckmark	&		\redcheckmark	&		\redcheckmark	&		\redcheckmark	&	\redcheckmark	&	\redcheckmark \\ 
\hline
12. & \textbf{TEROP}&	\ding{119} &	\ding{119}	&		\redcheckmark	&	\ding{119}	&		\redcheckmark	& 	\redcheckmark  \\ 
\hline
13. &  \textbf{STATIC}	&	\greencheckmark	&		\redcheckmark  	&	\greencheckmark	&	\ding{119}	&		\redcheckmark  &	\redcheckmark	 \\ 
\hline
14. &\textbf{ ENUM} &		\graycheckmark  	&		\redcheckmark  	&		\redcheckmark  	&		\ding{119}		&		\redcheckmark  	&	\ding{119}	\\ 
\Xhline{2\arrayrulewidth}
\multicolumn{7}{l} {\textbf{Flexible Invocations}} &	\\
\Xhline{2\arrayrulewidth}
1. & \textbf{EMPTY}	&		\redcheckmark  &		\redcheckmark  	&	\greencheckmark	&	\greencheckmark	&	\greencheckmark	& 	\redcheckmark	 \\ 
\hline
2. & \textbf{LOG}	&		\redcheckmark  &		\redcheckmark  	&		\redcheckmark  	&		\redcheckmark  	&		\redcheckmark  	&	\redcheckmark  	\\ 
\hline
3. & \textbf{CLIENT}&		\redcheckmark  	&		\redcheckmark  	&		\redcheckmark  	&		\redcheckmark  	&	\redcheckmark  	&	\redcheckmark  	\\ 
\hline
4. & \textbf{VAL}	&		\redcheckmark  &		\redcheckmark  	&		\redcheckmark  	&		\redcheckmark  	&		\redcheckmark  	&	\redcheckmark \\ 
\hline
5. & \textbf{HASH} &		\redcheckmark		&	\ding{119}	&	\ding{119}		&	\ding{119}		&	\ding{119}		&		\redcheckmark   \\ 
\hline
6. & \textbf{GETSUB} 	&	\redcheckmark  	&	\redcheckmark  	&	\redcheckmark  	&	\redcheckmark  	&	\redcheckmark  	&	\redcheckmark  	 \\ 
\hline
7. &\textbf{LEN/AUTH}	&	\redcheckmark  	&	\redcheckmark  	&	\redcheckmark  	&	\redcheckmark  	&	\redcheckmark  	&	\redcheckmark  	\\ 
\hline
8. & \textbf{GETPUB}	&  	\redcheckmark   &	\redcheckmark  	&	\redcheckmark  	&	\redcheckmark  	& \redcheckmark 	&	\redcheckmark  \\
\hline
9. &\textbf{STROP}	&	\redcheckmark  	&	\redcheckmark  	&	\redcheckmark  	&	\redcheckmark  	&	\redcheckmark  	&	\redcheckmark   \\
\Xhline{2\arrayrulewidth}
\end{tabularx}
\begin{flushleft}{
{\greencheckmark} = detected, {\ding{119}} = partially detected, {\redcheckmark} = undetected, \textbf{\graycheckmark} = undetected due to error\\
\textbf{METHOD*} = Multiple method calls, \textbf{STRBL*}=StringBuilder\\
}




\end{flushleft}
\label{tab:crypto-tools}
\end{table}

\myparagraph{Results} As seen in Table~\ref{tab:crypto-tools}, most tools struggle at detecting even the most basic 23 unnatural cases.
Only 5/23 cases were detected ({\greencheckmark}) by at least one tool.
No tool could reason about 12/23 cases, representing over 12,876/59,002 invocations, which highlights a significant gap in the detection capabilities of existing tools when dealing with real-world invocations which are even mildly unconventional. 
One welcome development is that instead of silently failing,  CogniCrypt declares that it cannot reason about the usage in 4 of these 12 cases (representing 7092/59,002 instances), \ie\ {\graycheckmark}.

\finding{Popular tools struggle to reason about basic forms of unnatural invocations uncovered in this study (n=23). No tool could reason about 12/23 cases, representing 12,876/59,002 invocations. In only 5/23 cases was there at least one tool that could reason about it. Finally, there was no case that every tool could detect.}

Finally, we marked the remaining 6/23 cases (representing 7,161/59,002 invocations) as {\em partially} detected (\ding{119}), as they represent the following situations where the tool cannot reason about the misuse but still reports a warning: the tool {\sf (1)} misidentifies a correct use in the invocation as a misuse, while failing to detect the actual misuse, or {\sf (2)} detects only one misuse in an invocation containing more than one, or, {\sf (3)} detects only some instances of a specific case, indicating hardcoded rules, and not the ability to detect a type of invocation.

\section{Discussion}
\label{sec:discussion}
Our characterization of crypto-API usage in the wild demonstrates that developers invoke crypto-APIs in highly diverse, complex, and non-intuitive ways in the wild.
Our experiments further demonstrate that popular detection tools cannot reason about many  prevalent, non-trivial, usage patterns.
Thus, this work motivates the re-thinking of security-analysis techniques to detect vulnerabilities in unnatural invocations, as we can no longer pretend they do not exist.
To this end, we distill the $17$ key findings into $4$ discussion themes that capture the challenges for vulnerability detection in this domain, and highlight actionable opportunities for future research. 

\subsection{Even {\em natural} cases need attention} 
\label{sec:simple-natural}
Tool designers often perceive the use of identifiers and string literals as a natural way for developers to use APIs such as \cipherget, in contrast to more complex usage that is deemed out of scope for detection.
However, we observe that even such seemingly natural patterns may exhibit unnaturalness in practice, which tools are unprepared for. 
For instance, identifiers or string literals may represent OIDs that point to vulnerable ciphers (\fnumber{2}), {\em which generally receive little attention from detection tools}. 

This gap between perceived naturalness and code in practice holds true for flexible API as well.
Conventional wisdom dictates that \checkServerTrusted method implementations that are empty are vulnerable, but those containing some code generally perform certificate validation checks.
Tools designed to detect misuse in these methods follow the same intuition~\cite{fahl2012eve}, and identify vulnerability with empty/sparse methods, thus ignoring methods with some code. 
However, we find that methods that score high on our complexity metric, \ie that contain significant code, exhibit unnatural and often vulnerable behavior, such as simply logging certificates, returning true via a method chain that does not perform any checks (\fnumber{14}), or only checking the expiration date (\fnumber{13}).

To summarize, we observe that broad assumptions by tool designers regarding syntactically conventional usage, \eg single String literals in \cipherget, or an non-empty \checkServerTrusted method, may not hold against real-world usage with unexpected semantics. 

\begin{takeaway}{Takeaway 1}%
Even if the consensus is that tools should only target conventional/natural misuse, we still need to re-evaluate our assumptions about what such natural expression entails.
\end{takeaway}%

\subsection{Evasion is Prevalent and Complex}
Prior work has discussed how developers may leverage complex patterns to evade vulnerability detectors~\cite{ami2022crypto}, and 
developers have described observations of evasion in practice~\cite{ami-fn-oakland24,mn25}.
Our analysis uncovered unique cases of potential evasion of not only automated tools, {\em but also manual code reviews}. 

Particularly, we found several cases where developers would present a benign argument in code, but actually use a vulnerable one (\fnumber{8}, \fnumber{10}). 
For instance, we found cases where a vulnerable argument (\eg\ \code{ECB} mode) was used to encrypt data, but was not visible in code, as it was encrypted using a secure algorithm (\ie\ \code{AES} in \code{GCM} mode) (\eg\ Listing~\ref{lst:decode}, and Listing~\ref{lst:fulldecrypt} in the Appendix). 
Similarly, we found code that uses XORs to change secure parameter values to vulnerable ones at runtime (\eg\ Listing~\ref{lst:cbctoecb} in the Appendix).
Further, we found several cases where developers would use String operations such as \code{charAt} to present \code{GCM} use in code, but transform it to \code{ECB} at runtime, as recent work has observed~\cite{mao+24} (\eg\ Listing~\ref{lst:charat}).
Such use would visually resemble the use of the secure algorithm for encryption, thereby evading both security tools as well as manual code reviews.

Beside these unambiguous attempts at evasion, we also found code with complex expression of vulnerable parameters, which may be potentially to evade automated tools.
For instance, we found the use of \code{StringBuffer} API and simple string transformations to combine arguments in non-intuitive ways (\eg appending "/EC" and "B/...", which essentially results in ECB mode) (\fnumber{7}), or through complex encoding and encryption (\fnumber{8}).
Further, we found evidence of developers hiding vulnerable parameters in native code (\fnumber{3}), \ie where the API itself would be called in Java, but the parameters would be fetched at runtime via a JNI call.

These findings are particularly concerning because in most cases of such evasive use, we observe that the eventually resolved API parameters often tend to be vulnerable. As we discuss in Ethical Considerations (Appendix~\ref{app:ethics}), we have reported all such usage as potential backdoors to Google, in addition to reporting them to app developers. 

\begin{takeaway}{Takeaway 2}%
We can no longer exempt evasive misuse from the design goals of automated vulnerability detection tools, as evasion is present, complex, and may indicate backdoors.
\end{takeaway}%

\subsection{What motivates unnatural usage?} 
One resounding question that may come up given our analysis results is: {\em why do developers implement such complex, non-intuitive, misuse?}
One rationale could be performance. 

This is particularly true in the case of mobile-IoT applications that need to interface with IoT devices and the cloud in realtime. 
Using a secure cipher such as \code{AES/GCM/NoPadding} may cost more runtime relative to a vulnerable option such as \code{AES/ECB/NoPadding}, motivating developers to use the latter. 
Alternately, developers may be motivated by tight deadlines, and may take the easy way out of tweaking invocations to finish the job, instead of avoiding or fixing vulnerabilities, as recent work has found~\cite{ami-fn-oakland24}.
That is, they may use vulnerable options such as empty \checkServerTrusted methods purely because they work, or write evasive code to get around a quality assurance in the organization, particularly if they are independent contractors with no stake in the product.
Developer forums such as StackOverflow certainly do not improve the status quo, \eg there are several examples of developers struggling to use \code{GCM} for encrypting and decrypting large files~\cite{stackx1,stackx2,stacko1}, with experts recommending \code{AES/ECB/NoPadding} and \code{AES/CBC/PKCS5Padding} because they ``eliminate the inconvenience''. 

To summarize, several practical factors, such as performance, deadlines, and a lack of ownership in the product, may incentivize developers to invoke crypto-APIs in convoluted, unnatural ways, often resulting in misuse and vulnerabilities. 
\begin{takeaway}{Takeaway 3}%
 As these factors incentivizing unnatural and evasive code are not expected to disappear, the resulting unnatural code will likely proliferate as well. This further motivates proactive re-design of existing analysis techniques to account for unnatural and/or evasive misuse.
\end{takeaway}%

\subsection{Toward Effective Detection Grounded in a Real-World Characterization}
\label{sec:discussion-detection}
A large-scale qualitative study  is the first step in developing an evidence-based understanding of real-world crypto-API usage, and the challenges in detecting it.
Particularly, we note that our findings would not have been possible using existing tools (\ie without the qualitative study), as existing tools fail to detect most basic types of invocations, let alone complex ones (see Section~\ref{sec:discussion-tools})  
This underscores a broader lesson -- that {\em understanding a problem in the wild is a prerequisite for building informed, effective, tools to detect it}.

The taxonomies and findings from this work lay the foundation for exactly this understanding in the context of crypto-API misuse. 
In doing so, this paper enables several short and long-term research opportunities for improving the state of the crypto-API misuse detection in practice, as described below.

\myparagraph{1. Helping security tools to systematically triage real-world crypto-API invocations} 
The taxonomies from this work are a precise classification of real-world crypto-API usage, accompanied by real code examples representative of each unique class.
Existing crypto-API misuse detectors will be able to leverage our taxonomies to significantly improve their outcomes.
At the very least, tools will be able to precisely declare what specific expressions of crypto-APIs from the taxonomies they cannot handle, using the code examples in the taxonomies to test if needed.
Such declaration will not only help tools systematically identify gaps and improve, but will also improve tool documentation, and enable tool users to adjust their expectations and understand tool output. 

Building on this, tools will be able to triage invocations at runtime, by matching invocations they encounter with classes in the taxonomy, using both   complexity scores and the distinctive arguments/variables that exemplify each class. 
Such triaging will help tools identify invocations they cannot reason about without significant analysis, and thus provide helpful information to the tool user for further analysis.

\myparagraph{2. Enabling benchmarks and evaluation frameworks grounded in real usage}
The results from our analysis and the taxonomies will provided a basis for testing crypto-API misuse detectors, and particularly when it comes to developing automated frameworks for the same.
For instance, frameworks such as MASC~\cite{ami2022crypto}, which develop synthetic variants of crypto-API misuse using mutation, will be able to directly adopt classes from our taxonomy (and accompanying code) as a base cases for mutation, as well as inspiration for developing new mutation operators. 
The mutants and resultant benchmark developed from such a framework will be highly relevant, as it will be grounded in systematically classified crypto-API usage in the wild.

\myparagraph{3. Enabling find-tuned AI support to complement SAST} As we see in this study, a major challenge for detectors is the interpretation of complex or unnatural usage. 
However, the detailed taxonomies of code with text descriptions produced in this work, in conjunction with the present developments in large language models (LLMs), exposes a significant research opportunity to address this problem.

To elaborate, we envision LLMs trained using few-shot learning, made possible by the real code examples and accompanying text descriptions of crypto-API usage classes from the taxonomies.
Such LLMs can then be used to make the detection of misuse in unnatural crypto-API invocations practical.
For instance, the LLMs could be used to resolve the effective arguments, parameters, or other identifying factors relevant to a specific class, which could then be checked to detect misuse.
Or, the LLMs could be used to simplify the invocation, \ie to summarize it further in a more simpler form that existing SAST tools can reason about, thereby pre-processing code for detection by SASTs. 
While it is unclear how well LLMs will be able to reason about crypto-API invocations in general, or how suitable they would be in detecting vulnerable use  (although a very recent evaluation shows promise~\cite{xia2024LLM,masood2024LLM}), the multi-modal ground truth data from this study will certainly give LLMs an edge in this problem space by infusing domain knowledge. 
\begin{takeaway}{Takeaway 4}%
 The ground-truth characterization of real-world crypto-API misuse from this work enables actionable opportunities for improving, benchmarking, and re-designing crypto-API detection techniques (\eg using LLMs).
\end{takeaway}%

\section{Threats to Validity}
\label{sec:threats}
We now describe the threats to the validity of our study.
\myparagraph{1. Manual Analysis, exploitability} Given that we seek to find the very unnatural variants of vulnerabilities that automated tools are incapable of detecting~\cite{ami2022crypto}, using automated tools or building a detector, was incompatible with our research goal.
Instead, our objective required us to meticulously construct the ground truth taxonomies, which may future research on building automated tools to detect such latent vulnerabilities.
That said, due to the manual approach, we do not claim completeness in terms of capturing all unnatural patterns. 
Further, we consider validating the exploitability of the misuse out of the scope of this work, as the focus is understanding the nature of vulnerable crypto-API invocations in order to improve their detection. 
That said, prior work shows the detection of well-known types of misuse generally resolves to true positives (\eg 24.7\% false alarms for \code{ECB} alarms from CryptoREX, as per Chen et al~\cite{chenndss2024}).
\myparagraph{2. Generalizability with respect to non-IoT apps} Our decision to focus on mobile-IoT apps was influenced by the {\em impact} from the critical attack surface they represent, and their likelihood of using crypto-APIs.
This decision resulted in a rich and large set of invocations for qualitative analysis, leading to significant results and findings.
That said, to explore if any new patterns would be found in an analysis of the broader Android app population, we extended our analysis to all non-IoT apps from Jin et al.'s dataset~\cite{jmk+22} that are now available on Google Play, \ie a sample of $409$ confirmed non-IoT Android apps.
We extracted $2,728$ crypto-API invocations ($2,451$ restrictive and $277$ flexible) from these apps.
We found that the distribution of complexity scores in non-IoT apps, for both the flexible and restrictive case, are strikingly similar to those seen for mobile-IoT apps in Figure~\ref{fig:restrictive} and Figure~\ref{fig:flexible} (see Appendix~\ref{app:generalizability-apps} for non-IoT app figures), except for the lack of empty \cipherget invocations (\score{-1}) in the non-IoT apps.
Moreover, our qualitative analysis of the $2,728$ invocations did not reveal any new patterns.
\myparagraph{3. Generalizability with respect to other crypto-APIs} Our results reflect the usage of two highly popular and well-studied crypto-APIs, \cipherget and \checkServerTrusted, which have also been historically well-represented in vulnerability research in this domain~\cite{ami2022crypto,jmk+22,pinningpradeep,hazhirpasand2020,nadi2016,swk+2022cambench,fahl2012eve,chenndss2024}. 
To explore generalizability to other crypto-APIs, we repeated our study on two additional APIs that are highly relevant in mobile application security: {\sf (1)} \secretkeyspec, which is used to handle the construction of cryptographic keys from byte arrays, and mostly falls under the {\em restrictive} invocation category, and {\sf (2)} \hostnameVerifier, from the {\em flexible} category. 
We extracted $10,568$ invocations of both APIs, of which $3,459$ were sampled for manual analysis.
On analyzing these invocations, we found no new patterns; instead, we found repeats of patterns that form our taxonomies.
The distributions of complexity scores for \secretkeyspec and \hostnameVerifier were also similar to those for \cipherget (\ie Figure~\ref{fig:restrictive}) and \checkServerTrusted (\ie Figure~\ref{fig:flexible}) respectively. 
Thus, while we do not claim generalizability, these additional insights demonstrate that characterization is representative of a dominant subset of interesting patterns of crypto-API invocation in the wild.
Detailed results from this additional analysis can be found in Appendix~\ref{app:generalizability-APIs}.

\section{Related Works}
\label{sec:related_work}
This paper presents the first study of the unnatural use of crypto-API in the wild. 
As a result, we discuss closely related work pertaining to SAST tools \& frameworks for finding crypto-API misuse~\cite{egele2013empirical,mobsf,fahl2012eve,kjk+2021cognicrypt,kruger2017cognicrypt,crylogger,rahaman2019cryptoguard,muslukhov2018,wickert2021,zcd+2019cryptorex}, existing benchmarks \& evaluation frameworks~\cite{ami2022crypto,clw+2024,rahaman2019cryptoguard,swk+2022cambench,sun2023,schlichtig2023} and qualitative studies~\cite{ami-fn-oakland24,hazhirpasand2020,nadi2016,kruger2023,oliveira2018,votipka2020,gorski2020,green2016}.

MalloDroid by Fahl et al.~\cite{fahl2012eve} was one of the first SASTs to evaluate Android apps for SSL/TLS-misuse vulnerabilities. 
Some of our findings resonate with MalloDroid, as even 12 years later, empty \checkServerTrusted methods remain extremely common and as shown in Table~\ref{tab:crypto-tools}, MalloDroid still performs significantly better than state-of-the-art crypto-detector tools. 
Egele \etal~\cite{egele2013empirical} developed a tool called CryptoLint which is based on six basic rules which Android apps should adhere to ensure indistinguishability under chosen plaintext attacks (IND-CPA). They found that $88\%$ of mobile apps violate at least one rule. 
Similar rules were employed by Zhang \etal~\cite{zcd+2019cryptorex} to design a tool called CRYPTOREX and applied the rules in the context of IoT firmwares, finding similar problems, \ie close to $25\%$ of IoT firmware violates at least one rule. 
In order to better support developers, Krüger \etal developed CogniCrypt using $23$ rules to ameliorate the problem with use of low level cryptographic primitives~\cite{kjk+2021cognicrypt}. Similarly, Cryptoguard tool and benchmark were designed based on $16$ rules~\cite{rahaman2019cryptoguard}.
These types of techniques used by state of the art crypto-detector tools may be effective against simple types of misuse, and even perform well on benchmarks from tool designers, such as CryptoAPI-Bench~\cite{rahaman2019cryptoguard} and CamBench~\cite{swk+2022cambench}.

Finally, some of our observed invocations bear significant resemblance to the synthetic mutants generated by MASC~\cite{ami2022crypto}. 
Particularly, the cases with string operations and concatenation observed in our analysis of restrictive crypto-API use are similar to the mutation operators built for MASC.
However, it is important to note that our work is focused on finding relevant real world examples in the wild, and is the first qualitative characterization of unnatural misuse in this space.

\section{Conclusion}
\label{sec:conclusion}
This paper performed the first large-scale qualitative study (n=$5,704$ API invocations) of odd/unnatural use of cryptographic-APIs in the wild. 
The resultant taxonomies capture this characterization in 96 unique classes of invocation, many of them non-trivial.
Our 17 findings effectively resolve the initial contention posed in the paper, and show that a diverse spectrum of unnatural usage is prevalent in the wild, and should be within the scope of work of crypto-API misuse detection tools.
More importantly, we show that existing detectors do not detect even most of the basic cases of unnatural uses from the taxonomies, thereby motivating the re-design of tools with a focus on such hard-to-detect vulnerabilities.
The results, findings and discussion from this paper emphasize the practical challenges for vulnerability detection in the face of unnatural API invocation, and describe actionable opportunities to leverage the insights and artifacts from this study for improving vulnerability detection in this domain.

\section*{Ethical Considerations}
\label{app:ethics}
\myparagraph{Stakeholders} We identified two primary stakeholders that may be impacted by our work, these include the application developers and the end user. 
In order to reduce harm to users and app developers, have followed a responsible vulnerability disclosure process, described later in this section.
Moreover, it is also important to note that we do not disclose application names in the paper, as again, the goal is not to study specific apps, but the nature of certain misuse.

\myparagraph{Responsible Disclosure}
Following the principle of beneficence, we conducted responsible disclosure prior to publicizing our work. 
We have reported all the misuse found in our qualitative study (i.e., analysis of $5,704$ crypto-API invocations) to the appropriate vendors/app developers, and in some cases, also to Google.
To elaborate, we submitted reports containing one or more discovered misuse to $241$ app developers.
In the case of $161$ applications, where we found the expression of the misuse to be highly evasive, we also reported the misuse as a potential crypto backdoor to Google, in order to protect users. 
We noted that at the time of our reporting, some of the applications with evasive use cases had already been taken down from Google Play, including three applications that hid vulnerable parameters in native method calls and one that used base64.

\myparagraph{Potential Harm Mitigation}
It is important to highlight that this research did not experiment with live systems, as we only statically analyzed code obtained from mobile apps. However, we automatically downloaded the APKs corresponding to our list of mobile apps  (obtained from Jin et al.~\cite{jmk+22}) from Google Play.
To prevent any adverse impact on Google Play, we did not run multiple scripts concurrently, and only only ran a single downloader, with a $5$ secs sleep timer after every APK searched/downloaded. 
We did not collect developer information such as name, affiliation or/and emails. 
We also did not collect user reviews and/or users PII associated with apps from Google Play. 
Finally, we followed a responsible vulnerability disclosure process, which mitigates harm to users (in the form of undisclosed vulnerabilities) and app developers (in the form of harm to reputation from irresponsible disclosure of vulnerabilities).

\myparagraph{Decision} Our research aims to measure the unnaturalness of cryptographic-API (mis)use in the wild. This work uncovers several critical cryptographic-API misuse which may have eluded research and industry security tools. We carefully weighed the potential benefits of the research against any possible harm, and concluded that our research provides a net positive to the community.

\bibliographystyle{plain}
{\footnotesize
	\bibliography{references}
}

\cleardoublepage
\appendix

\begin{table*}[h!]
	\scriptsize
	\centering
	\def\arraystretch{1.5}
	\caption{\small Restrictive Taxonomy Labels }
	\begin{tabularx}{\textwidth}{p{1in}|p{1.0in}|p{4.4in}}
		\Xhline{2\arrayrulewidth}
		\textbf{ Label} &\textbf{Label Name} & \textbf{Description of Labels}\\
		\Xhline{2\arrayrulewidth}
		\hline
		\textbf{STROP} &  String Operation &This refers to the use of any of the string methods such as \texttt{charAt}, \texttt{format}, \texttt{toString} etc. as depicted in Android developers  documentation~\cite{androidstring}. We associate every use of such method with this label. Listing~\ref{lst:stroa}\\
		\hline
		\textbf{TEROP} &  Ternary Operator  &This refers to the use of the conditional operator that accepts three operands. This mimics the if-else statement such that based on a conditional (first operand), any of the two other operand could be the return value. Listing~\ref{lst:tera}\\
		\hline
		\textbf{ENUM}&  Enum Type Constant** & Enum Type is a class used to store predefined constant in code such as various cipher algorithms, mode, padding etc. Enum type offers more functions such as use with switch case statements, type and value safety etc. Listing~\ref{lst:ena}\\
		\hline
		\textbf{ID} &  Identifier & Identifiers refers to unique variable name particularly assigned to cipher algorithm, mode, and/or padding.Listing~\ref{lst:ida} \\
		\hline
		\textbf{THIS}&  Reference Current Object & Referencing the current object refers to using the ``this'' keyword to refer to the current object instance variable or method. Listing~\ref{lst:refa}\\
		\hline
		\textbf{METHOD} &  Method Call &  Used to indicate the presence of a  \textit{function} call that returns the cipher algorithm., mode, and/or padding. Listing~\ref{lst:meta}\\
		\hline
		\textbf{STATIC} & Static Field Constant  & Similar to enum type, it is used to store predefined values, however, unlike enum type  it does not offer additional functions.Listing~\ref{lst:staa}\\
		\hline
		\textbf{NATIVE} & Native Method Call  & This refers to the accessing methods or variables implemented in C/C++ and stored in native libraries. Listing~\ref{lst:nava}\\
		\hline
		\textbf{BAS64}&  Base64-Decoder & This refers to decoding byte data encoded using Base64 encoding scheme. Listing~\ref{lst:basa}\\
		\hline
		\textbf{STRBUF} &  String Buffer  & This refers to the use of ``append'' method to add strings in a sequence. StringBuilder is broadly similar to StringBuffer however, StringBuilder is faster under most implementation. There are many more methods associated with this class and should be considered for future investigations. Listing~\ref{lst:sba}\\
		\hline
		\textbf{CONCT} & Concatenation  & This refers to the use of the addition operator ``+'' to sequentially append string i.e. cipher algorithm, mode and padding, to each other. Listing~\ref{lst:cona}\\
		\hline
		\textbf{SEPRT}&  Separator  & This refers to ``/'' which typically separates the cipher algorithm, mode and padding, used as a standalone value either as a string, identifier, static field class etc.  Listing~\ref{lst:sepa}\\
		\hline
		\textbf{OID}&  Object Identifier  & These are numeric strings used to refer to cipher algorithms. Listing~\ref{lst:oida}\\
		\textbf{STRING}&  String    & This refers a series of characters i.e. string literals that refers to the cipher algorithm, mode and/or padding. Listing~\ref{lst:stra}\\
		\hline
		\textbf{EMPTY} &  Empty  & This refers to an empty arguments \ie cipher algorithms \code{Cipher.getInstance()}. Listing~\ref{lst:empa} \\
		
		\Xhline{2\arrayrulewidth}
	\end{tabularx}
	\label{tab:res_tax}
\end{table*}

\begin{table*}[h!]
	\scriptsize
	\centering
	\def\arraystretch{1.5}
	\caption{\small Flexible Taxonomy Labels }
	\begin{tabularx}{\textwidth}{p{1in}|p{1.0in}|p{4.4in}}
		\Xhline{2\arrayrulewidth}
		\textbf{ Label} &\textbf{Label Details} & \textbf{Description of Labels}\\
		\Xhline{2\arrayrulewidth}
		\hline
		\textbf{ABS} &  Abstract Method &  This is a method declared without a method body i.e. curly braces -- ``\{\}'' and terminted with a semicolon.\\
		\hline
		\textbf{NATIVD} &  Native Method Declaration   & This refers to checkServerTrusted method defined in native code. It is distinctively identified by the use of keyword ``native''.\\
		\hline
		\textbf{ILL}&  Throwing Illegal Argument Exception & This refers to throwing the Illegal Argument Exception, if certificate chain and auth type values are null or zero-length.\\
		\hline
		\textbf{CEXP} &  Throwing Certificate Exception & This refers to throwing the Certificate Exception, if certificate chain is not trusted by the TrustManager. \\
		\hline
		\textbf{NEXP}&  Throwing Null and Asserion Error &This refers to unconditionally throwing Null or Assertion Error only in the method body.\\
		\hline
		\textbf{LOG} &  Logging Certificates  & This refers logging certificates in the method body of checkServerTrusted method. \\
		\hline
		\textbf{NATIVE} & Native Method Call  &This refers to the accessing methods or variables implemented in C/C++ and stored in native libraries \\
		\hline
		\textbf{METHOD} & Method Call   & Used to indicate the presence of a  \textit{function} call that returns the verifies the certificate\\
		\hline
		\textbf{THIS}&  Reference Current Object  & Referencing the current object is using the ``this'' keyword to refer to the current object instance variable or method. \\
		\hline
		\textbf{ISTRST} &  TrustStrategy $\rightarrow$ isTrusted Method  & This refers to the use of an interface particularly a method named ``isTrusted'' which determines whether the certificate chain can be trusted without consulting the trust manager configured in the SSL context~\cite{istrusted}.\\
		\hline
		\textbf{VAL} & Checking Certificate Validity  & This refers to one of the x509Certificate method used to check if certificate is currently valid i.e. not expired~\cite{androidx509cert}\\
		\hline
		\textbf{LEN}&  Checking Certificate Chain Length  &This refers checking if certificate is zero-length. \\
		\hline
		\textbf{NULL}&  Non-Null Value Check  & This refers checking if certificate material is non null. \\
		\hline
		\textbf{LIST}&  Using List Methods    & Lists is often used to handle certificate materials, this refers to the use of list methods to i.e. add, remove etc while handling certificate.
		\\
		\hline
		\textbf{CERFAC} &  Certificate Factory  & This is used to generate certificate object and initializes it with the data read from an input stream~\cite{certfactgen}.\\
		\hline
		\textbf{VER} &  Verify Method&  This is the use of one of X509Certificate methods which verifies that a certificate was signed using a private key that corresponds to public key specified as input~\cite{androidx509cert}\\
		\hline
		\textbf{STROP} &  String Operation  & This refers to the use of any of the string methods such as \texttt{contains}, \texttt{equals}, etc.  \\
		\hline
		\textbf{AUTH}&  AuthType Value Check &  This refers checking Auth Type string for type of algorithm used, checking for zero-length and/or null value.
		\\
		\hline
		\textbf{ENCOD} &  getEncoded Method & This refers to the use ``getEncoded'' method associated with an abstract class for managing different types of certificates with common functionality i.e. X.509, PGP etc~\cite{certall}\\
		\hline
		\textbf{ARR}&  Using Array Methods & This refers to the use of arrays methods in handling or verifying certificates. \\
		\hline
		\textbf{BIGINT} & BigInteger  & This is the use of BigInteger.for managing large integer numbers. It is often use to handle certificate materials such as public keys. \\
		\hline
		\textbf{TMFAC} & TrustManager Factory &This is acts a factory for trustmanagers which are responsible for handling trust materials used in establishing trust. \\
		\hline
		\textbf{GETPUB} & Certificate.getPublicKey  & It is often used to return the public key of the root certificate. Used possibly for certificate validation purposes. \\
		\hline
		\textbf{GETISR}& Certificate.getIssuerDN (Deprecated)  & This is used to return the issuer distinguished value which is the entity that signed and issued the certificate. This method is deprecated and replaced with ``getIssuerX500Principal''~\cite{androidx509cert}. \\
		\hline
		\textbf{GETSUB} &  Certificate.getISubjectDN (Deprecated)  & This method returns the subject (subject distinguished name) value from the certificate. This method is also deprecated and replaced with ``getSubjectX500Principal''. \\
		\hline
		\textbf{HASH}&  Hash Algorithm  & This refers to the use of hash algorithms such as SHA1 and SHA256 for certificate validation purposes.\\
		\hline
		\textbf{CERPAT} &  CertPathValidator  & This is used for validating certificate chains or path. This is used to return a CertPathValidator object that implements the specified algorithm i.e. ``PKIX''\\
		\hline
		\textbf{PKIX} & PKIXParameters  & This is used to populates the set of most-trusted CAs from the trusted certificate entries contained in the specified KeyStore.It is often used to disable the revocation enable flag to enable faster validation.\\
		\hline
		\textbf{CLIENT}&  Calling checkClientTrusted   & This refers calling an empty checkClientTrusted from checkServerTrusted method for certificate validation. \\
		\hline
		\textbf{EMPTY}&  Empty Method Body  & This refers to a \checkServerTrusted method with empty body i.e. no certificate validation logic implemented.\\

		\Xhline{2\arrayrulewidth}
	\end{tabularx}
	\label{tab:flex_tax}
\end{table*}

\section{Non-IoT Application Analysis}
\label{app:generalizability-apps}

\begin{figure}[h!]
	\def\arraystretch{1.5}
	\centering
	\includegraphics[width=3.4in]{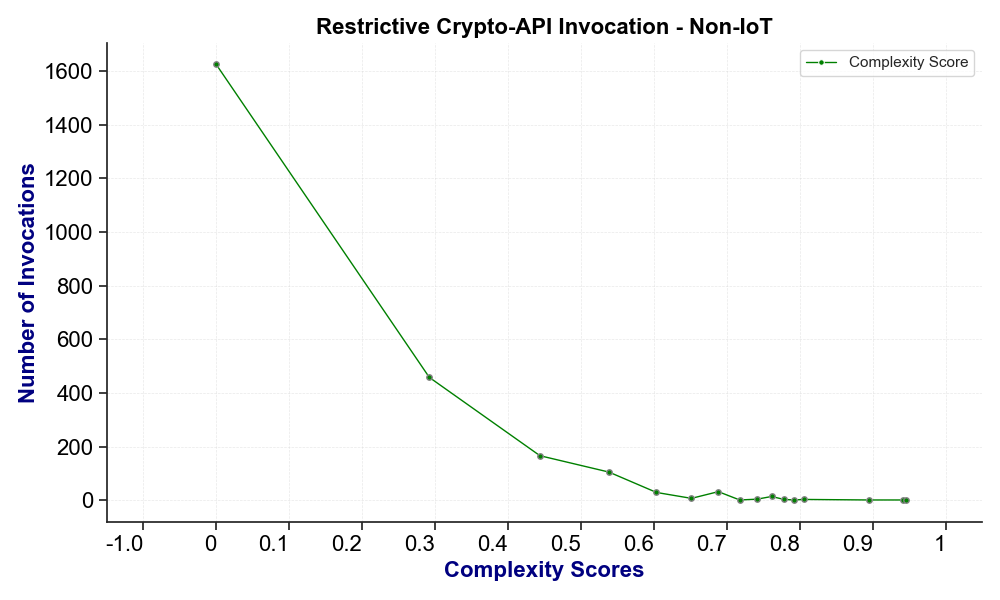}
	\caption{{\small {Complexity Score Distribution for Restrictive Invocation for Non-IoT Apps.}}}
	\label{fig:restrictive-non-iot}
\end{figure}

\begin{figure}[h!]
	\def\arraystretch{1.5}
	\centering
	\includegraphics[width=3.4in]{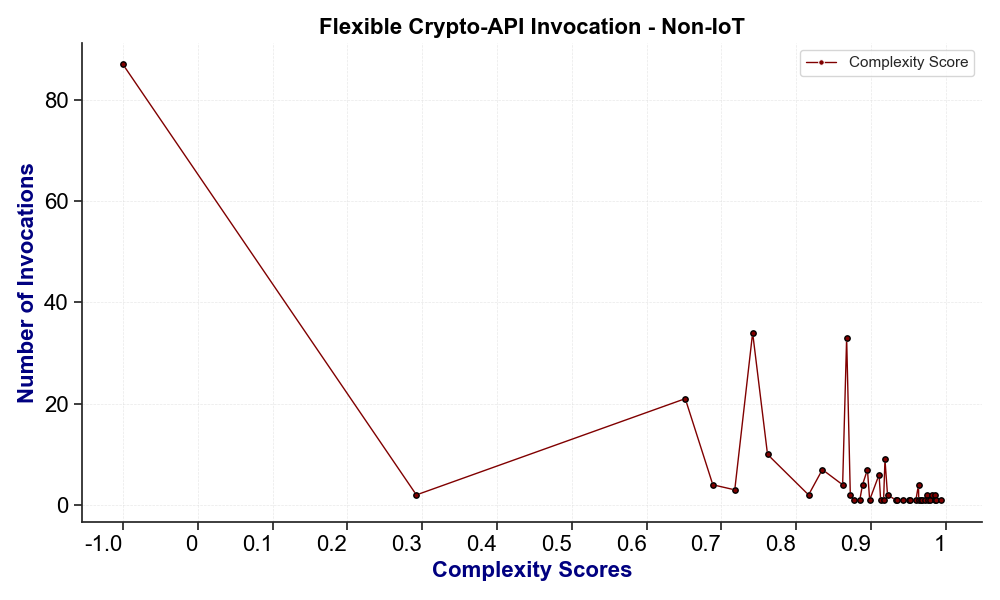}
	{\footnotesize \textbf{Note:} Score $-1$ has $87$ number of invocations and is not in the graph.}
	\caption{{\small {Complexity Score Distribution for Flexible Invocation for Non-IoT Apps.}}}
	\label{fig:flexible-non-iot}
\end{figure}

We decompiled $409$ mobile apps (non-IoT) and found $2,451$ Cipher.getInstance restrictive invocations and $277$ checkServerTrusted flexible invocations. Figure~\ref{fig:restrictive-non-iot} \&~\ref{fig:flexible-non-iot} show the distribution of complexity scores vs number of invocations for restrictive and flexible invocations for non-IoT apps respectively. Both figures can be observed to be highly similar to their respective counterpart for mobile-IoT apps which indicates that the trends observed in mobile-IoT apps is similar to that in non-IoT apps. 
For restrictive invocations, \score{0} had $1,625$ invocations with the use of identifiers only being $462$ and use of string only being $1,163$ invocations. \score{0.2923} had $535$ invocations, \score{0.4439} had $89$ invocations, \score{0.5385} had $105$, and the rest from \score{0.6 - 0.9363} having a total of $97$ invocations with 0.9363 being the maximum score. 
For flexible invocations, our analysis found 277 invocations for checkServerTrusted with $31.4$\% ($87$/$277$) having a \score{-1}i.e. implementing an empty Trust manager. This is similar to the $34.2$\% of invocations implementing an empty Trust manager for mobile-IoT apps. \score{0} accounts for $3.61$\% ($10$/$277$) of the dataset which consists of abstract methods ($4$/$277$) and native method declarations (6/277). This is very similar to \score{0} for mobile-IoT apps. 

We did not observe any new patterns on qualitative analysis of the $2,728$ invocations from non-IoT apps. 
Instead, we observed two additional salient examples of two key types in the restrictive taxonomy (Figure~\ref{fig:tax_res}), namely the \code{ID} and \code{METHOD} type, which we discuss as follows:

\myparagraph{Evasive Example, Identifier use} One key finding for identifier use as shown in Listing~\ref{lst:cbctoecb} is assigning it to a method call. 
This method call takes a cipher algorithm as an argument i.e. ``AES/CBC/PKCS5Padding''. 
The key idea here is to get the byte value of the string `C', `B', `C', and perform bit-wise XOR with the byte value of `6',`1',`1' \ie `C' byte value is XOR-ed with `6' and so on. 
This operation produces `ECB' when converted to string. We developed minimal working examples using code found in live applications. We found 6 instances of this used for encryption and decryption across multiple applications.

\myparagraph{Example of Method calls} As shown in Listing~\ref{lst:byte}, method calls were used to extracts specific byte values from a hard-coded list of bytes, offsets the byte values by a specific number, and appends each character to a StringBuilder before converting the output to string. 
This type of code led to 3 instances of using ``AES'' and 2 instances of using ``RSA/ECB/PKCS1PADDING'' cipher algorithms. 
We found 5 instances of this use in the same mobile application.

\section{Generalizability with respect to other Crypto-APIs -- Additional Details}
\label{app:generalizability-APIs}

We find similar trends as regards complexity scores for \secretkeyspec (see online appendix~\cite{onlineappendix}), except for missing \score{0} and \score{-1}. \score{0} is missing because the baseline use case for this API requires two arguments, unlike \cipherget, which has an optional argument to specify a provider. We find the use of hardcoded keys (Listing~\ref{lst:key1}), use of base64 (Listing~\ref{lst:base64key}), method call, stringBuffer, use of identifiers to store hardcoded keys, use of enum, concatenation, which reveals that developers often use similar techniques we observed in our analysis of \cipherget. 
The only difference we observed is in one case, where the developers stored the generated key material in a file, which is not an essential aspect of the crypto-API invocation itself. 

The trends observed for \hostnameVerifier are similar to those in \checkServerTrusted. We found empty implementations or those that always return true, allowing all hostnames. We observed more complex implementations of \hostnameVerifier that call another method that returns true (i.e., \fnumber{14}), which essentially turns hostname verification off. 
Most complex implementations of \hostnameVerifier often call other methods to handle individual components of the verification. \eg one method checks the host name format (IPV4, IPV6, URL), and another component uses elements of the X509Certificate, such as its getSubjectAltName, and compares that with the hostname or IP address.

\section{Restrictive Invocations -- Code Snippets}

\noindent
\begin{minipage}{\linewidth}
\begin{lstlisting}[frame=tb,caption={String Operation}, label={lst:stroa},language=java,basicstyle=\fontencoding{T1}\fontfamily{lmtt}\bfseries\scriptsize]
Cipher.getInstance(String.format("%s/%s/%s", "RSA", "ECB", "PKCS1Padding"));
\end{lstlisting}
\end{minipage}

\noindent
\begin{minipage}{\linewidth}
        \begin{lstlisting}[frame=tb,caption={{\small The use of nested ternary operators in \ciphergetinstance}}, label={lst:slash},language=java]
Cipher.getInstance(Intrinsics.areEqual(this. sharedPreferences.getString("cipher.used", Build.VERSION.SDK_INT >= 23 ? "M" : "PREM"), "M") ? "RSA/ECB/OAEPWithSHA-1AndMGF1Padding" : "RSA/ECB/PKCS1Padding")
        \end{lstlisting}
\end{minipage}

\noindent
\begin{minipage}{\linewidth}
\begin{lstlisting}[frame=tb,caption={Ternary Operator}, label={lst:tera},language=java,basicstyle=\fontencoding{T1}\fontfamily{lmtt}\bfseries\scriptsize]
Cipher.getInstance(z ? "AES/GCM/NoPadding" : "AES/CBC/PKCS5Padding");
\end{lstlisting}
\end{minipage}

\noindent
\begin{minipage}{\linewidth}
\begin{lstlisting}[frame=tb,caption={Enum Type}, label={lst:ena},language=java,basicstyle=\fontencoding{T1}\fontfamily{lmtt}\bfseries\scriptsize]
Cipher.getInstance(aVar.f13865d);
\end{lstlisting}
\end{minipage}

\noindent
\begin{minipage}{\linewidth}
\begin{lstlisting}[frame=tb,caption={Identifier}, label={lst:ida},language=java,basicstyle=\fontencoding{T1}\fontfamily{lmtt}\bfseries\scriptsize]
Cipher.getInstance(nativeGetString); 
\end{lstlisting}
\end{minipage}

\noindent
\begin{minipage}{\linewidth}
\begin{lstlisting}[frame=tb,caption={Referencing Current Object.}, label={lst:refa},language=java,basicstyle=\fontencoding{T1}\fontfamily{lmtt}\bfseries\scriptsize]
Cipher.getInstance(this.mTransformation);
\end{lstlisting}
\end{minipage}

\noindent
\begin{minipage}{\linewidth}
\begin{lstlisting}[frame=tb,caption={Method Call}, label={lst:meta},language=java,basicstyle=\fontencoding{T1}\fontfamily{lmtt}\bfseries\scriptsize]
Cipher.getInstance(getCipherAlgorithm());
\end{lstlisting}
\end{minipage}

\noindent
\begin{minipage}{\linewidth}
\begin{lstlisting}[frame=tb,caption={Static Final Class.}, label={lst:staa},language=java,basicstyle=\fontencoding{T1}\fontfamily{lmtt}\bfseries\scriptsize]
Cipher.getInstance(SecurityConstants.AES_MODE);
\end{lstlisting}
\end{minipage}

\noindent
\begin{minipage}{\linewidth}
\begin{lstlisting}[frame=tb,caption={Native Method Call.}, label={lst:nava},language=java,basicstyle=\fontencoding{T1}\fontfamily{lmtt}\bfseries\scriptsize]
Cipher.getInstance(requestTransformation(1));
\end{lstlisting}
\end{minipage}

\noindent
\begin{minipage}{\linewidth}
\begin{lstlisting}[frame=tb,caption={Base64 for Decoder}, label={lst:basa},language=java,basicstyle=\fontencoding{T1}\fontfamily{lmtt}\bfseries\scriptsize]
Cipher.getInstance(new String(Base64.decode("REVTL0NCQy9QS0NTNVBhZGRpbmc=", 2)));
\end{lstlisting}
\end{minipage}

\noindent
\begin{minipage}{\linewidth}
\begin{lstlisting}[frame=tb,caption={String Buffer/String Builder}, label={lst:sba},language=java,basicstyle=\fontencoding{T1}\fontfamily{lmtt}\bfseries\scriptsize]
Cipher.getInstance(stringBuffer.toString());
\end{lstlisting}
\end{minipage}

\noindent
\begin{minipage}{\linewidth}
\begin{lstlisting}[frame=tb,caption={Concatenation}, label={lst:cona},language=java,basicstyle=\fontencoding{T1}\fontfamily{lmtt}\bfseries\scriptsize]
Cipher.getInstance("AES/ECB/" + padding);
\end{lstlisting}
\end{minipage}

\noindent
\begin{minipage}{\linewidth}
\begin{lstlisting}[frame=tb,caption={Separator.}, label={lst:sepa},language=java,basicstyle=\fontencoding{T1}\fontfamily{lmtt}\bfseries\scriptsize]
Cipher.getInstance(str5 + RDMConstants.SLASH + str2 + RDMConstants.SLASH + str4, provider);
\end{lstlisting}
\end{minipage}

\noindent
\begin{minipage}{\linewidth}
\begin{lstlisting}[frame=tb,caption={Object Identifier.}, label={lst:oida},language=java,basicstyle=\fontencoding{T1}\fontfamily{lmtt}\bfseries\scriptsize]
Cipher.getInstance("1.2.840.113549.3.2");
\end{lstlisting}
\end{minipage}

\noindent
\begin{minipage}{\linewidth}
\begin{lstlisting}[frame=tb,caption={String.}, label={lst:stra},language=java,basicstyle=\fontencoding{T1}\fontfamily{lmtt}\bfseries\scriptsize]
Cipher.getInstance("Blowfish");
\end{lstlisting}
\end{minipage}

\noindent
\begin{minipage}{\linewidth}
\begin{lstlisting}[frame=tb,caption={Empty Argument}, label={lst:empa},language=java,basicstyle=\fontencoding{T1}\fontfamily{lmtt}\bfseries\scriptsize]
AESCipher.getInstance();
\end{lstlisting}
\end{minipage}

\noindent
\begin{minipage}{\linewidth}
\begin{lstlisting}[frame=tb,caption={XOR and Base64 with up to 4 method calls}, label={lst:xor1},numbers=left,numbersep=5pt,
numberstyle=\scriptsize\bfseries,language=java,basicstyle=\fontencoding{T1}\fontfamily{lmtt}\bfseries\scriptsize]
Cipher cipher = Cipher.getInstance(cipherAlgorithm);
----------
cipherAlgorithm = 
Utils.base64DecodeAndXor("Iio+ASgjKE4/ZSIjXDMOCUoCDww=");
-----------
static String base64DecodeAndXor(String str) {
 return xorMessage(new String(Base64.getDecoder().decode(str)));
--------------
Note: xorMessage in Line 7 calls two more methods not shown
cipherAlgorithm = "AES/CBC/PKCS5Padding"
\end{lstlisting}
\end{minipage}

\noindent
\begin{minipage}{\linewidth}
\begin{lstlisting}[frame=tb,caption={Extracting AES from byte array}, label={lst:byte},numbers=left,numbersep=5pt,
numberstyle=\scriptsize\bfseries,language=java,basicstyle=\fontencoding{T1}\fontfamily{lmtt}\bfseries\scriptsize]
Cipher cipher = Cipher.getInstance(m7713c());
-------------------------
public static String m7713c() {
int[] iArr = {55, 49, 59, 54, 47, 45, 36, 43, 41};
StringBuilder sb = new StringBuilder();
for (int i2 = 8; i2 < 9 && i2 >= 0; i2 -= 3) {
 sb.append(Character.toChars(iArr[i2] + 24));
}
return sb.toString();
\end{lstlisting}
\end{minipage}

\noindent
\begin{minipage}{\linewidth}
\begin{lstlisting}[frame=tb,caption={Base64 Encoded Key},label={lst:base64key},numbers=left,numbersep=5pt,numberstyle=\scriptsize\bfseries,language=java,basicstyle=\fontencoding{T1}\fontfamily{lmtt}\bfseries\scriptsize]
new SecretKeySpec(Base64.decode("oik6PdDdMnOXemTbwvMn9de/h9 lFnfBaCWbGMMZqqoSaQaqUOqjVGm5NqsmjcBI1x+sS9ugjB55HEJWR iFXYFw==", 2), "HmacSHA256")
\end{lstlisting}
\end{minipage}

\noindent
\begin{minipage}{\linewidth}
\begin{lstlisting}[frame=tb,caption={Direct Hard Coded Key}, label={lst:key1},language=java,basicstyle=\fontencoding{T1}\fontfamily{lmtt}\bfseries\scriptsize]
new SecretKeySpec("oejkdirztefhnvscxhdmdzedfotuabje".get Bytes("UTF-8"), "AES");
\end{lstlisting}
\end{minipage}

\noindent
\begin{minipage}{\linewidth}
\begin{lstlisting}[frame=tb,caption={SHA1 and getEncoded to check hard coded root cert.}, label={lst:sha1getenc},language=java,basicstyle=\fontencoding{T1}\fontfamily{lmtt}\bfseries\scriptsize]
public void checkServerTrusted(X509Certificate[] x509CertificateArr, String str2) throws CertificateException {
try {
if(AdjustBridgeUtil.byte2HexFormatted(MessageDigest .getInstance("SHA1").digest(x509CertificateArr[0] .getEncoded())) .equalsIgnoreCase("7BCFF44099A35BC093BB48C5A6B9A5
16CDFDA0D1")) {
return;
}
\end{lstlisting}
\end{minipage}

\noindent
\begin{minipage}{\linewidth}
\begin{lstlisting}[frame=tb,caption={Using GCM mode with Hard Coded Key to conceal use of ECB mode}, label={lst:fulldecrypt},language=java,basicstyle=\fontencoding{T1}\fontfamily{lmtt}\bfseries\scriptsize]
private static final String KEY_AES = "AES";  
private static final String KEY_CIPHER = "AES/GCM/NoPadding";  
public static final String KEY_GCM = "OGEseetime201800";  
private static String TAG = "AES2Utils";

algorithmStr_encode = "32Bi2A5oaH61xilScou92x9faAiO0SOBXmb0X/wqAijapt8K"
String AES_ECB_PADDING_encode = "32Bi2A5oaH6r4jpgI7+10Rwi6u+aWTgnrWUjLeHbiJK5";

Cipher.getInstance(AesGcmUtils.decode(algorithmStr_encode));
Cipher.getInstance(decode(AES_ECB_PADDING_encode);

--------------------------------------
public static String decode(String str) {  
 try {  
  return str.isEmpty() ? "" : new String(decrypt(getbase64ToBytes(str), KEY_GCM)); //Calls the function below with Base64.getDecode().decode  
 } catch (Exception e) {  
  e.printStackTrace();  
 return ;  
}  
}

-------------------------------
public static byte[] decrypt(byte[] bArr, String str) {  
 try {  
  SecretKeySpec secretKeySpec = new SecretKeySpec(getKey(KEY_GCM), KEY_AES);  
  byte[] bytes = str.getBytes(StandardCharsets.UTF_8);  
  Cipher cipher = Cipher.getInstance(KEY_CIPHER);  
  cipher.init(2, secretKeySpec, new GCMParameterSpec(128, bytes));  
 return cipher.doFinal(bArr);  
 } catch (Exception e) {  
  e.printStackTrace();  
 return new byte[0];  
}  
}

\end{lstlisting}
\end{minipage}

\noindent
\begin{minipage}{\linewidth}
\begin{lstlisting}[frame=tb,caption={Use of XOR to change CBC to ECB}, label={lst:cbctoecb},numbers=left,numbersep=5pt,
numberstyle=\scriptsize\bfseries,language=java,basicstyle=\fontencoding{T1}\fontfamily{lmtt}\bfseries\scriptsize]
Qhi = Qhi("AES/CBC/PKCS5Padding");
Cipher.getInstance(Qhi);
-------------
public static String Qhi(String str) {
 int[] iArr = new int[str.length()];
 iArr[4] = 6;
 iArr[5] = 1;
 iArr[6] = 1;
 return new String(Qhi(str.getBytes(), iArr));}
-------------
public static byte[] Qhi(byte[] bArr, int[] iArr) {
 if (bArr == null || bArr.length == 0 || iArr == null || iArr.length == 0) {
  return bArr;}
 byte[] bArr2 = new byte[bArr.length];
 for (int i = 0; i < bArr.length; i++) {
  bArr2[i] = (byte) (bArr[i] ^ iArr[i % iArr.length]);}
 return bArr2; }
\end{lstlisting}
\end{minipage}

\end{document}